%% file: root.tex
\pdfoutput=1
 \documentclass[final,5p,times,twocolumn,authoryear]{elsarticle}

\usepackage{amssymb}
\usepackage{amsmath}
\usepackage[version=4]{mhchem} 
\usepackage{tikz} 
    \usetikzlibrary{patterns}
    \usetikzlibrary{decorations.pathmorphing}
    \usetikzlibrary{decorations.markings}
    \usetikzlibrary{arrows.meta,bending}
\usepackage{subcaption} 
\usepackage{physics} 

\journal{Arxiv}

\begin{document}

\begin{frontmatter}




\title{Dynamical Simulation Model of the Pyro-Process in Cement Clinker Production }

\author[DTUComp,FLS]{Jan Lorenz Svensen} 
\author[FLS]{Wilson Ricardo Leal da Silva} 
\author[DTUComp]{Zhanhao Zhang}
\author[DTUComp]{Steen Hørsholt}
\author[DTUComp]{John Bagterp J\o rgensen\corref{cor1}} \ead{jbjo@dtu.dk}
\affiliation[DTUComp]{organization={Department of Applied Mathematics and Computer Science, Technical University of Denmark},
            city={Kgs. Lyngby},
            postcode={2800}, 
            country={Denmark}}
\affiliation[FLS]{organization={FLSmidth Cement A/S},
            city={Valby},
            postcode={2500}, 
            country={Denmark}}            

\cortext[cor1]{Corresponding author}


\begin{abstract}
This study presents a dynamic simulation model for the pyro-process of clinker production in cement plants. 
The study aims to construct a simulation model capable of replicating the real-world dynamics of the pyro-process to facilitate research into the improvements of operation, i.e., the development of alternative strategies for reducing \ce{CO2} emissions and ensuring clinker quality, production, and lowering fuel consumption.
The presented model is an index-1 differential-algebraic equation (DAE) model based on first engineering principles and modular approaches. Using a systematic approach, the model is described on a detailed level that integrates geometric aspects, thermo-physical properties, transport phenomena, stoichiometry and kinetics, mass and energy balances, and algebraic volume and energy conservations. 
By manually calibrating the model to a steady-state reference, we provide dynamic simulation results that match the expected reference performance and the expected dynamic behavior from the industrial practices.

\end{abstract}




\end{frontmatter}



\input{Introduction}

\input{Model}
\input{Simulations}

\input{Conclusion}

\section*{Acknowledgement}
This work is part of an Industrial PostDoc research (Project Number: 2053-00012B), the NEWCEMENT project of the INNO-CCUS partnership, and the Dynflex project of the MissionGreenFuels partnership (Project Number: 1150-00001B) granted by Innovation Fund Denmark and the Ecoclay project funded by Danish Energy Technology Development and Demonstration Program (EUDP) under the Danish Energy Agency (Project Agreement No. 64021-7009).
Also, the authors would like to thank Javier and Dinesh for their valuable contribution on practical insights related to cement production and process control).
%

\bibliographystyle{elsarticle-harv} 
\bibliography{biblio}




\end{document}

%% file: Introduction.tex
\section{Introduction}\label{sec:Introduction}
Cement is a mixed product, consisting primarily of cement clinker with small additions of gypsum.
On a global basis, cement manufacturing corresponds to 8\% of the world's \ce{CO2} emissions \citep{CO2Techreport}. The production of cement clinker is the main source of the \ce{CO2} emissions.
Thus, reducing the \ce{CO2} emissions from clinker production is of industrial interest \citep{chatterjee2018cement}. 

In clinker production, \ce{CO2} emissions come from the combustion of fuel and the calcination of calcium carbonate (limestone). The production is a pyro-process, where heat is generated by fuel combustion to facilitate calcination and other chemical reactions necessary to form quality clinker.
\ce{CO2} reduction can thus be achieved by 1) changing the heat source to a green option such as electricity, and 2) 
partially replacing clinker in the cement composition. The second option requires conservation of the cement properties to be viable, such as 28-day strength. 
The current approach to the second option focuses on lowering the content of clinker in the cement (the clinker ratio) by utilizing Supplementary Cementitious Materials (SCM) as a substitute.
The remaining clinker ratio is used to ensure the quality of the cement, i.e., its chemical and physical properties are within existing technical standards.

Thus to maximize the substitution of the clinkers with alternative materials, it is necessary to know the quality of the remaining clinker, i.e., the clinker composition and variability.

\begin{figure*}[ht]
    \centering
    \includegraphics[width=\textwidth,trim={0cm 40cm 0cm 0cm},clip]{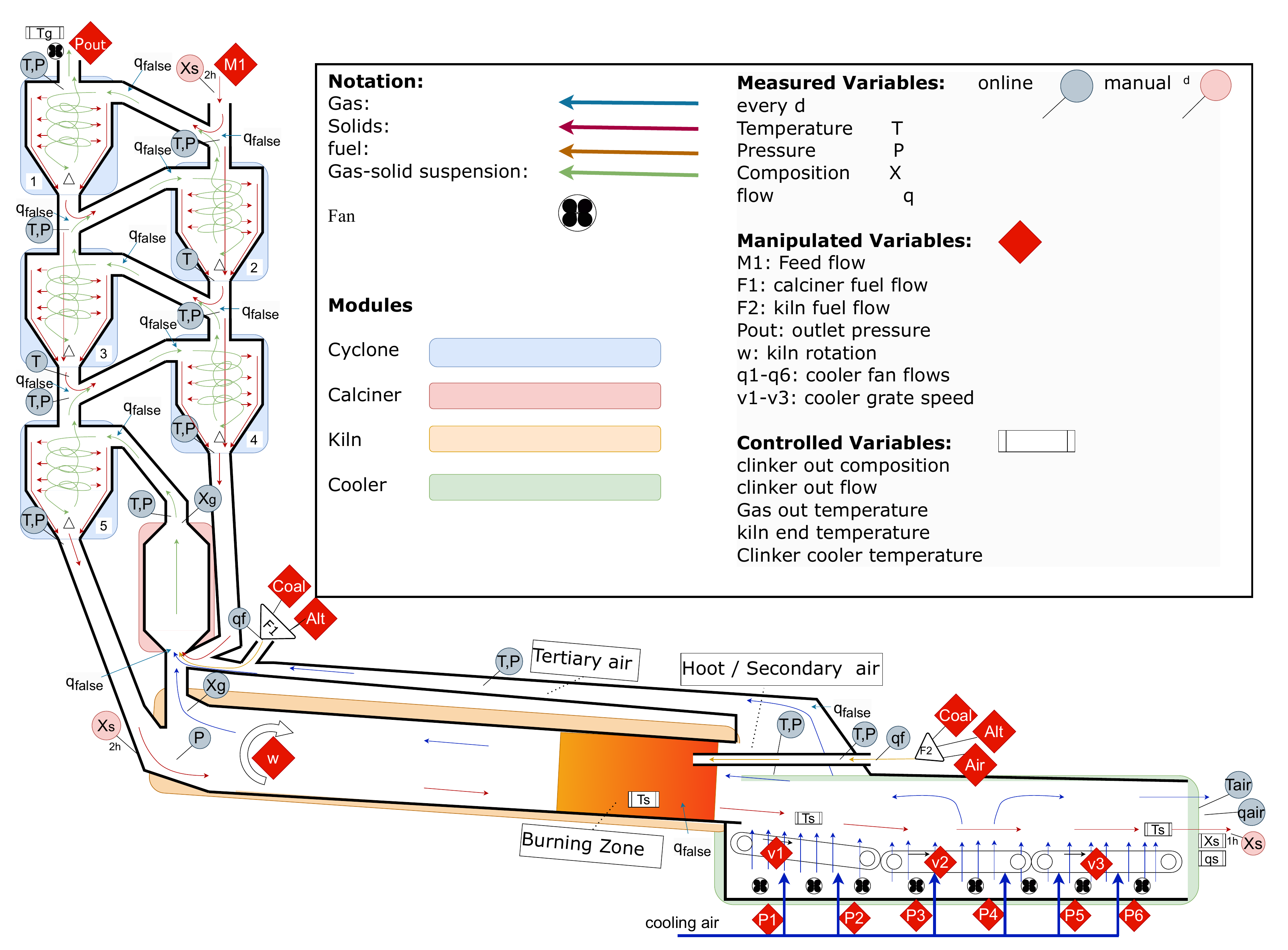}
    \caption{A conceptual layout of the pyro-process of clinker production. The flow patterns of feed material, gas content, and their mixture are indicated by directional arrows.}
    \label{fig:process}
\end{figure*}

To achieve reliable knowledge of the clinker quality and transition towards zero \ce{CO2}-emission cement plants, digitalization, control, and optimization methods are necessary tools.
The development of such tools requires digital twins for dynamic simulation of the cement plant of interest. Fig. \ref{fig:process} illustrates the layout of a typical pyro-process for clinker production.
The process consists of a set of cyclones, a kiln, a cooler, a calciner, and risers (i.e., ducts connecting each part).

Literature on practical industrial operation and design practices 
 includes \cite{bhatty2010innovations},\cite{chatterjee2018cement}, and \cite{bye1999portland}.
A few full-process models are available in the literature, though a dynamic simulation model describing the full pyro-process in detail and suitable for control is not available. 

\cite{Kolip2019} provides a steady-state model for the full pyro-process of a 5-stage single-string system.
The model includes the gas and solid phases, described by mass and energy balances for the accumulated compounds, with the compound ratios as variables to be computed.
\cite{MUJUMDAR2007} presents a simulation model with a dynamic calciner model and steady-state models for the kiln, the cooler, and the cyclone parts. The calciner model applied discrete particle models for coal particles and raw meal particles, and continuous phase models for the overall gas balance and species mass fraction.

Additional models on the individual process parts are available in the literature. The literature on cyclones includes the work of Barth and Muschelknautz \citep{Hoffmann07} describing steady-state models of cyclones based on geometric analysis and the work of \cite{Mohamadali2023} and \cite{Park2020}, who applied CFD modeling with finite volumes to describe cyclones.
The literature on calciner models includes a CFD model by \cite{Kahawalage2017} and a 1D dynamic Eulerian model by \cite{ILIUTA2002805}.
Kiln models in the literature include 
the steady-state energy and mass balance model based on temperatures and mass fractions by \cite{Mastorakos1999CFDPF}, 
The steady-state model of mass-energy balances with bed variation by \cite{Mujumdar2006},
the steady energy balance model by \cite{Hanein2017},
the dynamic model with chemistry and flame model by \cite{Spang},
The dynamic model with momentum balance without reactions by \cite{LIU2016}, and the dynamic material with chemistry and steady-state gas model by \cite{GINSBERG2011}.
 The cooler models in the literature include the dynamic 1D energy balance model by \cite{METZGER1983491}, 
 the steady-state 2D energy balance model by \cite{CUI20171297}, and the steady-state 1D mass-2D energy balance model by \cite{MUJUMDAR2007}.

To extend the literature, we provide a mathematical model of the full pyro-process for dynamic simulations. Using a rigorous approach to thermo-physical properties and kinetics, the model is formulated using a systematic modeling methodology based on first engineering principles. 
The resulting model is a system of index-1 differential algebraic equations (DAEs), integrating the mass and energy balances with thermo-physical properties, transport phenomena, and stoichiometry and kinetics. 

This paper collects and builds on our prior papers on the individual parts of the pyro-process: the calciner \citep{Svensen2024Calciner}, the kiln \citep{Svensen2024Kiln}, the cooler \citep{Svensen2024Cooler}, and the cyclones \citep{Svensen2024Cyclone}. The extensions include integration of the modules, overall process tuning, and boundary flows, e.g., false air.

\subsection{Paper outline}
This paper is organized as follows: section \ref{sec:Calciner} presents the pyro-process in cement clinker production, section \ref{sec:FullModel} the mathematical model, Section \ref{sec:SimulationResults} the process simulations,  section \ref{sec:Discussion} discussion of the results, and section \ref{sec:Conclusion} the conclusions.

\section{The pyro-process}
\label{sec:Calciner}
In the pyro-process of clinker production, the material feed is heated to facilitate 1) calcination of the feed and 2) the formation of clinker by further heating. The heat source is typically the combustion of fuel.

The pyro-process consists of several divisions:
\vspace{-2mm}
\begin{itemize}
     \setlength\itemsep{0.0em}
    \item A preheating tower consisting of cyclones connected by riser pipes and a calciner.  Here, the material feed is preheated and most of the calcination is facilitated.
    \item A kiln, where the temperature is high enough for clinker formation.
    \item A cooler, where the formed clinker is cooled rapidly to stabilize it, and heat is recuperated towards the kiln and calciner.
\end{itemize}
\vspace{-3mm}
The fuel is injected in the calciner and kiln parts.

In this paper, we consider the case of a clinker production line with a single-string 5-stage preheating tower, a rotary kiln, and a grate-belt cooler.
Fig. \ref{fig:process} shows the full pyro-process considered, illustrating how each module interconnects and which manipulated variables are available and where.
The model is formulated in terms of the modules: cyclones, calciner and risers, kiln, and cooler:
\begin{itemize}\setlength\itemsep{0.0em}
    \item The cyclones are chambers designed to separate suspended solids from gas in the preheating tower by inducing a swirling motion in the gas.
    \item The calciner and risers are pipes of varying shapes. Upon entrance, inflows are assumed uniformly mixed. The calciner has an injection of fuel, which facilitates the release of heat through combustion.
    \item The rotary kiln is a long rotating tube with a slight tilt. The feed material moves down the tube while the gas flows upward. The rotation induces a cascading motion in the feed to ensure even heating of the materials. The gas is heated by the combustion of the injected fuels.
    \item The grate-belt cooler is a conveyor belt with cooling fans beneath it.
The fans blow ambient air upward through the belt to cool the clinker.
The belt transports the hot clinker from the kiln over the cooling fans as the final step in the pyro-process.
The heated air in the cooler is recirculated into the kiln and calciner to recuperate the heat.
\end{itemize}

%% file: Model.tex
\section{Dynamic simulation model}\label{sec:FullModel}
We consider different phases in each module: solid phase ($_s$), gas phase ($_g$), solid-gas mixture phase ($_m$), refractory phase ($_r$), and/or wall phase ($_w$). The cyclones, risers, and calciner modules consider the mixture, refractory, and wall phases. The kiln considers solid, gas, and refractory phases. The cooler considers the solid and gas phases.


\subsection{Model approach and notation}

We apply a finite-volume approach to describe each module in segments of equal length along their axial axis, $\Delta y$ or $\Delta z$. An exception is the cyclones, these are considered lumped models of a single volume.
We define the molar concentration vector, $C$, as mole per segment volume $V_\Delta(k)$.
We apply the following assumptions: \begin{itemize}
    \item All gasses are ideal.
    \item The dynamics are along the axial dimensions, with the modules being homogeneous along the non-axial dimensions.
    \item No dust or trapped gas exists in the kiln and cooler, i.e., no material exchange between the phases.
    \item No backflows happen unless stated.
    \item Only the main clinker formation reactions and basic fuel reactions are included.
\end{itemize}

We use the standard cement chemist notation for the following compounds: \ce{(CaO)_2SiO_2} as \ce{C_2S}, \ce{(CaO)_3SiO_2} as \ce{C_3S}, \ce{(CaO)_3Al_2O_3} as \ce{C_3A}, and \ce{(CaO)_4} \ce{(Al_2O_3)(Fe_2O_3)} as \ce{C_4AF}, where C = \ce{CaO},   A = \ce{Al_2O_3}, S = \ce{SiO_2}, and F = \ce{Fe_2O_3}.
 
\subsection{Mass and energy balances}
The differential part of each module consists of the mass balances of each compound and the energy balances of each phase.
Thus, in the following, $N$ is molar flux, $\tilde{H}$ is enthalpy flux, $Q$ is heat transfer, and $J$ is phase transition heat transfer.
\subsubsection{Cyclones}
Fig. \ref{fig:process} shows that the cyclones have one inflow of solid-gas mixture, one outflow of separated solids, and one outflow of exiting solid-gas mixture.
Thus, the mass balances of compound $i$ for the suspended solids and gas phase is
\begin{subequations}
\begin{align}
    \partial_t C_{s,i} &= \frac{A_{in}N_{s,in,i}-A_{x}N_{s,x,i}-A_{sep}N_{sep,i}}{V_\Delta}  + R_{s,i},\\
    \partial_t C_{g,i} &= \frac{A_{in}N_{g,in,i}-A_{x}N_{g,x,i}}{V_\Delta}  + R_{g,i}.
\end{align}
\end{subequations}
The energy balances are given by
\begin{subequations}
\begin{align}
    \partial_t \hat{U}_m &= \frac{1}{V_\Delta}(\Delta\tilde{H}_s + \Delta\tilde{H}_g  - Q^{cv}_{mr} - Q^{rad}_{mr}),\\
    & \Delta\tilde{H}_s = A_{in}\tilde{H}_{s,in}-A_{x}\tilde{H}_{s,x}-A_{sep}\tilde{H}_{s,sep},\\
    & \Delta\tilde{H}_g = A_{in}\tilde{H}_{g,in}-A_{x}\tilde{H}_{g,x},\\
    \partial_t \hat{U}_r &= \frac{1}{V_r}(Q^{cv}_{mr}+Q^{rad}_{mr} - Q^{cv}_{rw}),\\
    \partial_t \hat{U}_w &= \frac{1}{V_w}(Q^{cv}_{rw} - Q^{cv}_{we} - Q^{rad}_{we}).
\end{align}
\end{subequations}

\subsubsection{Calciner and risers}
For the calciner and risers, the inflows enter at the bottom and exit at the top. Thus, the mass balances for the solid phase and the gas phase are
\begin{subequations}
\begin{align}
    \partial_t {C}_{s,i} &= -\partial_yN_{s,i} + R_{s,i},\\
    \partial_t {C}_{g,i} &= -\partial_yN_{g,i} + R_{g,i}.
\end{align}
\end{subequations}
The energy balances are given by
\begin{subequations}
\label{eq:EnergyBalances}
\begin{align}
    \partial_t\hat{U}_m &= -\partial_y(\Tilde{H}_m+ \Tilde{Q}_m) - \frac{Q^{rad}_{mr} + Q^{cv}_{mr}}{V_{\Delta}}, \\
    \partial_t\hat{U}_r &= - \partial_y\Tilde{Q}_r + \frac{Q^{rad}_{mr} + Q^{cv}_{mr}}{V_{r}} - \frac{Q^{cv}_{rw}}{V_{r}},\\
    \partial_t\hat{U}_w &= - \partial_y\Tilde{Q}_w + \frac{Q^{cv}_{rw}}{V_{w}} - \frac{Q^{rad}_{we} + Q^{cv}_{we}}{V_{w}} .
\end{align}
\end{subequations}
\subsubsection{Kiln}
The flows in the kiln are counter-flows. The solid flows down the kiln, and the gas flows up the kiln along its length. The mass balances read
\begin{subequations}
\begin{align}
    \partial_t C_{s,i} &= -\partial_zN_{s,i} + R_{s,i},\\
    \partial_t C_{g,i} &= -\partial_zN_{g,i} + R_{g,i}.
\end{align}
\end{subequations}
The energy balances are formulated as
{\small
\begin{subequations}
\begin{align}
    \partial_t\hat{U}_s &= -\partial_z(\Tilde{H}_s + \Tilde{Q}_s) + \frac{Q^{rad}_{gs}+Q^{rad}_{ws} + Q^{cv}_{gs} + Q^{cv}_{ws}}{V_{\Delta}} - J_{sg},\\
    \partial_t\hat{U}_g &= -\partial_z(\Tilde{H}_g+\Tilde{Q}_g) - \frac{Q^{rad}_{gs}+Q^{rad}_{gw}+Q^{cv}_{gs}+Q^{cv}_{gw}}{V_{\Delta}} + J_{sg},\\
    \partial_t\hat{U}_w &= -\partial_z\Tilde{Q}_w+ \frac{Q^{rad}_{gw}- Q^{rad}_{ws} + Q^{cv}_{gw} - Q^{cv}_{ws}}{V_{\Delta w}}.
\end{align}
\end{subequations}
}

\subsubsection{Cooler}
The flows in the cooler are cross-flows. The gas flows from beneath and up, while the solid flow is from side to side. The mass balances are given as
\begin{subequations}\label{eq:mass} 
    \begin{align}
        \partial_tC_{s,i} &= -\nabla \vdot N_{s,i} + R_{s,i}, \qquad \nabla = [\partial_y,\partial_z]^T,\\
        \partial_tC_{g,i} &= -\nabla \vdot N_{g,i} + R_{g,i}.
    \end{align}    
\end{subequations}     
The energy balances are formulated as
\begin{subequations}
\begin{align}
    \partial_t\hat{U}_s &= -\nabla \vdot (\Tilde{H}_s + \Tilde{Q}_s)  - \hat{Q}^{rad}_{sg} - \hat{Q}^{cv}_{sg} - J_{sg}, \\
    \partial_t\hat{U}_g &= -\nabla \vdot( \Tilde{H}_g + \Tilde{Q}_g ) + \hat{Q}^{rad}_{sg}+\hat{Q}^{cv}_{sg} + J_{sg}.
\end{align}
\end{subequations}

\subsection{Algebraic equations for volume and internal energy}
In each segment of each module, the pressure is defined by the conservation of the total segment volume, by the algebraic relation
\begin{align}
    \hat{V}_{g} + \hat{V}_{s} = \hat{V}_{\Delta} = 1.
\end{align}
The temperatures of each phase are defined by the conservation of energy, through the phase energy densities:
\begin{subequations}
    \begin{align}
    \hat{U}_g &= \hat{H}_g - P\hat{V}_g, \quad &\hat{U}_s &= \hat{H}_s, \quad &\hat{U}_r &= \hat{H}_r,\label{eq:EnergyAlgebra}\\
     \hat{U}_m &= \hat{H}_s + \hat{H}_g - P\hat{V}_g, \quad &\hat{U}_w &= \hat{H}_w.
\end{align}
\end{subequations}
For the kiln module, the algebraic relations include \eqref{eq:geoAlg} to compute the fill angle $\theta$.

\begin{table}[t]
    \centering
    \caption{Material properties of the solid phase.}
    \begin{footnotesize}  
    \begin{tabular}{c|ccc}\hline
           &\shortstack{Thermal\\ Conductivity} & Density & \shortstack{Molar \\Mass}\\ \hline\\[-1em]
            &$k_s$ & $\rho_s$ & $M_s$\\ \hline\\[-1em]
         Units    & $\frac{\text{W}}{\text{Km}}$ & $\frac{\text{g}}{\text{cm}^3}$ & $\frac{\text{g}}{\text{mol}}$ \\[-1em]\\ \hline
         \ce{CaCO_3} &  2.248$^a$& 2.71$^b$  &100.09$^b$\\ 
         \ce{CaO}     & 30.1$^c$ &  3.34$^b$ &56.08$^b$\\ 
         \ce{SiO_2}   &  1.4$^{a,c}$& 2.65$^b$  &60.09$^b$\\ 
         \ce{Al_2O_3} &  12-38.5$^c$ 36$^a$& 3.99$^b$  &101.96$^b$\\ 
         \ce{Fe_2O_3} &  0.3-0.37$^c$& 5.25$^b$  &159.69$^b$\\ \hline
         \ce{C2S}    &  3.45$\pm$0.2$^d$& 3.31$^d$  &$172.24^g$\\ 
         \ce{C3S}    & 3.35$\pm$0.3$^d$ & 3.13$^d$ & 228.32$^b$\\ 
         \ce{C3A}    &  3.74$\pm$0.2$^e$& 3.04$^b$ & 270.19$^b$\\ 
         \ce{C4AF}   &  3.17$\pm$0.2$^e$& 3.7-3.9$^f$  &$485.97^g$ \\ \hline        
    \end{tabular}   
    \end{footnotesize}
    
    \footnotesize{$^a$\cite{Perry}, $^b$ \cite{CRC2022}, $^c$ \cite{Ichim2018}, $^d$ \cite{PhysRevApplied}, $^e$ \cite{Du2021}, $^f$ \cite{bye1999portland}, $^g$ Computed from the above results.} 
    \label{tab:Data-Coeff-solid}
\end{table}
\begin{table}[t]
    \centering
    \caption{Material properties of the gas phase.}
    \def\arraystretch{1.5}
    \begin{footnotesize}        
    \begin{tabular}{c|cccccc}\hline
           &\multicolumn{2}{c}{\shortstack{Thermal\\ Conductivity$^a$}} & \shortstack{Molar\\ Mass$^a$} & \multicolumn{2}{c}{Viscosity$^a$} & \shortstack{Diffusion\\ Volume$^b$}\\ \hline
        &\multicolumn{2}{c}{$k_g$} & $M_g$ & \multicolumn{2}{c}{$\mu_g$} & $V_\Sigma$\\ \hline
        Units   & \multicolumn{2}{c}{$\frac{10^{-3}\text{W}}{\text{Km}}$} & $\frac{\text{g}}{\text{mol}}$ & \multicolumn{2}{c}{$\mu$Pa s} & cm$^3$\\ \hline
        Temp. & 300 K& 1000K&-&300 K& 1000K&-
        \\ \hline
         \ce{CO_2}    & 16.77 & 70.78 & 44.01  & 15.0 & 41.18 & 16.3\\
         \ce{N_2} & 25.97&  65.36  &28.014 & 17.89 & 41.54 & 18.5\\
         
         \ce{O_2}  & 26.49&  71.55 &  31.998 & 20.65 & 49.12&  16.3\\
         
         \ce{Ar}  & 17.84&43.58& 39.948& 22.74 &55.69&  16.2\\
         
         \ce{CO}  & 25&  43.2$^c$ & 28.010& 17.8&  29.1 &  18\\
         
         \ce{C_{sus}} & - & -& 12.011 & - & -&  15.9\\
         
         \ce{H_2O} & 609.50&  95.877 &18.015 & 853.74 &37.615&   13.1\\
         
         \ce{H_2} & 193.1 & 459.7 &2.016 & 8.938 & 20.73& 6.12\\\hline
    \end{tabular}       
    \end{footnotesize}

    \footnotesize{$^a$ \cite{CRC2022}, $^b$ \cite{Poling2001Book}, $^c$ value at T=600K.}
    \label{tab:Data-Coeff-gas}
\end{table}

\subsection{Thermo-physical models}
Each module is described by the same thermo-physical models for enthalpy, $H(T,P,n)$, volume, $V(T,P,n)$, heat capacity, $c(T)$, density, $\rho(C,\hat{V})$, viscosity, $\mu(T,C)$, and thermal conductivity, $k(T)$, specific to each phase.
Tables \ref{tab:Data-Coeff-solid} and \ref{tab:Data-Coeff-gas} show the material data of each compound.
\subsubsection{Enthalpy}
The thermo-physical description for the enthalpy $H$ is 
 \begin{align}
    H(T,P,n) &= \sum_i n_i \left(\Delta_f H^\circ_{i}(T_0,P_0) + \int^T_{T_0}c_{p,i}(\tau)d\tau \right).
\end{align}
$\Delta_f H_{i}(T_0,P_0)$ is the formation enthalpy at conditions $(T_0,P_0)$ and  $c_p$ the molar heat capacity of each compound.
As the model is homogeneous of order 1 in the mole vector, $n$, the phase $i$ enthalpy density, $\hat{H}_i$, and enthalpy flux, $\tilde{H}_i$, can be defined as
\begin{align}
    \hat{H}_i &= H(T_i,P,C_i),\quad
    \tilde{H_i} = H(T_i,P,N_i).
\end{align}
Table \ref{tab:forma} contains the formation enthalpy of each compound.
\begin{table}[t]
    \centering
    \caption{Formation enthalpies, $\Delta_fH^\circ$.}
    \begin{footnotesize}   
    \begin{tabular}{ccccccccc}\hline
          units&\ce{CaCO3}$^a$ &\ce{CaO}$^a$     & \ce{SiO2}$^a$   &  \ce{Al2O3}$^a$ &\ce{Fe2O3}$^a$\\ 
        $\frac{\text{kJ}}{\text{mol}}$ &- 1207.6 & -634.9 & -910.7 &-1675.7&	-824.2\\ \hline
        
        units& \ce{C2S}$^b$ &\ce{C3S}$^b$ & \ce{C3A}$^b$&\ce{C4AF}$^b$\\
        $\frac{\text{kJ}}{\text{mol}}$  & -2053.1&-2704.1&-3602.5 & -4998.6\\ \hline
        
          units&\ce{CO_2}$^a$ & \ce{N_2}$^a$ & \ce{O_2}(g)  & \ce{Ar}$^a$\\
         $\frac{\text{kJ}}{\text{mol}}$  & - 393.5& 0& 0& 0\\ \hline
        
          units& \ce{CO}$^a$  & \ce{C_{sus}}$^a$ &\ce{H2O}$^a$ &\ce{H2}$^a$\\
         $\frac{\text{kJ}}{\text{mol}}$  & -110.5& 0& -241.8& 0\\ \hline
    \end{tabular}\\
    $^a$ Data from \cite{CRC2022}, $^b$ Computed from \cite{Mujumdar2006} and Hess' law.
    \end{footnotesize}
    \label{tab:forma}
\end{table}
\subsubsection{Volume}
The formulations of the volume models are phase-dependent. The solid and gas phase formulations are
 \begin{align}
    V_s(T,P,n) &= \sum_i n_i \left( \frac{M_i}{\rho_{b,i}} \right), \quad
    V_g(T,P,n) =  \sum_{i} n_i \left( \frac{RT}{P} \right).
\end{align}
$M$ is the molar mass and $\rho_b$ the bulk density of the compound.
As with the enthalpy, the solid and gas phase volume densities can be defined as
\begin{align}
    \hat{V}_s = V_s(T,P,C_s),\quad  \hat{V}_g = V_g(T,P,C_g)
\end{align}
Since the model dynamics are formulated by the concentrations $C$, the solid and gas volumes of each segment can be obtained from their densities:
\begin{align}
     V_s = \hat{V}_sV_{\Delta}(k), \quad V_g = \hat{V}_gV_{\Delta}(k),.
\end{align}

\subsubsection{Heat capacity} 
The heat capacity of phase $j$, $c_j$, is given by
\begin{align}   
c_{j}(T,n) &= \sum_in_{j,i}c_{p,i}(T).
\end{align}
$c_p$ is the molar heat capacity. Table \ref{tab:heatCap} shows the coefficients for the molar heat capacity for the formula
\begin{align}
    c_p(T) = C_0 + C_1 T + C_2 T^2 + C_{-2}T^{-2} + C_{-\frac{1}{2}}T^{-\frac{1}{2}}.
\end{align}
The heat capacity of the solid-gas mixture phase is given by
\begin{align}
    c_{m} = c_{s} + c_{g}.
\end{align}

\begin{table}[t]
    \centering
    \caption{Molar heat capacities.}
    \begin{scriptsize}
    \begin{tabular}{c|c c  cc  c| c}\hline
            & $C_0$ & $C_1$ & $C_2$ &  $C_{-2}$& $C_{-\frac{1}{2}}$ &\shortstack{Temperature\\ range}\\ \hline
         Units & $\frac{\text{J}}{\text{mol}\cdot \text{K}}$& $\frac{10^{-3}\text{J}}{\text{mol}\cdot \text{K}^2}$&$ \frac{10^{-5}\text{J}}{\text{mol}\cdot \text{K}^3}$ & $ \frac{10^{6}\text{J K}}{\text{mol}}$& $ \frac{\text{J}}{\text{mol}\cdot \text{K}^{0.5}}$&K\\ & & & & & &\\[-1em]\hline   
         & & & & & &\\[-0.8em]
         \ce{CaCO3}$^a$ & -184.79 & 0.32 & -0.13 & -3.69& 3883.5& 298 - 750\\[0.2pt]
         \ce{CaO}$^b$     &  71.69& -3.08  & 0.22   &0 &0 & 200 - 1800\\[0.2pt] 
         \ce{SiO_2}$^b$   & 58.91 &  5.02 & 0 & 0& 0&844 - 1800\\ [0.2pt]
         \ce{Al_2O_3}$^b$ &  233.004&-19.59   &0.94   &0 &0 & 200 - 1800\\[0.2pt] 
         \ce{Fe_2O_3}$^c$ & 103.9  & 0 & 0  & 0&0 &-\\ \hline
         & & & & & &\\[-0.8em]
         \ce{C2S}$^b$     &  199.6& 0  &0   &0 &0 &1650 - 1800\\ [0.2pt] 
         \ce{C3S}$^b$     &  333.92&  -2.33&  0 &0 &0 &200 - 1800\\ [0.2pt]
         \ce{C3A}$^b$    & 260.58  & 9.58/2 & 0  &0 &0 &298 - 1800\\ [0.2pt]
         \ce{C4AF}$^b$  &  374.43& 36.4 & 0  & 0&0 & 298 - 1863\\ \hline
         & & & & & &\\[-0.8em]
         \ce{CO_2}$^c$    & 25.98 &43.61 &-1.49  & 0& 0& 298 - 1500\\[0.2pt]
         \ce{N_2}$^c$ & 27.31&5.19 &-1.55e-04  &0 &0 &298 - 1500\\ [0.2pt]
         \ce{O_2}$^c$ & 25.82&12.63 &-0.36   &0 &0 &298 - 1100\\ [0.2pt]
         \ce{Ar}$^c$ & 20.79 & 0 & 0   & 0&0 &298 - 1500\\ [0.2pt]
         \ce{CO}$^c$ & 26.87& 6.94  & -0.08   & 0& 0& 298 - 1500\\[0.2pt] 
         \ce{C_{sus}}$^c$& -0.45& 35.53 & -1.31  &0 &0 &298 - 1500\\ [0.2pt]
         \ce{H_2O}$^c$&30.89 & 7.86  &0.25  &0 & 0&298 - 1300\\ [0.2pt]
         \ce{H_2}$^c$&   28.95& -0.58&  0.19 & 0& 0& 298 - 1500\\ \hline 
    \end{tabular}        
    \end{scriptsize}
    \footnotesize{ $^a$ \cite{Jacobs1981}, $^b$ \cite{HANEIN2020106043}, $^c$ \cite{CRC2022}. }
    \label{tab:heatCap}
\end{table}

\subsubsection{Density} 
The particle density of phase $j$, $\rho_j$, is given by
\begin{align}
    \rho_j=\rho(C_j,\hat{V}_j) = \frac{1}{\hat{V}_j}\sum_iM_iC_{j,i}.
\end{align}
 The volume density, $\hat{V}_j$, scales it to the actual volume of the particles.
The segment density of phase $j$, $\bar\rho_j$, is given by
\begin{align}
    \bar\rho_j=\bar\rho(C_j) = \sum_iM_iC_{j,i}.
\end{align}
The density of the solid and gas phase mixture, $\rho_m$, is given by
\begin{align}
    \bar\rho_m =\bar\rho_s + \bar\rho_g.
\end{align}

For the kiln and cooler modules, the solid phase rests on the process equipment in a bulk. Thus for these modules, the density of the bulk is utilized. The bulk density depends on the particle density, $\rho$, by
\begin{align}
    \rho_b = (1-\phi_p)\rho,
\end{align}
where $\phi_p$ is the porosity of the bulk.
The porosity depends on the bulk composition, e.q. sulfur content. Assuming the clinkers are ideal spheres of equal size, then according to \cite{hales_adams_bauer_dang_harrison_hoang_kaliszyk_magron_mclaughlin_nguyen_etal._2017}, the range of porosity is 
\begin{align}
    1-\frac{\pi}{3\sqrt{2}} \leq \phi_p \leq 1.
\end{align}
In the literature, \cite{CUI20171297} proposed a constant porosity of 0.4 for the bulk in the cooler.

\subsubsection{Viscosity and thermal conductivity}
The viscosity, $\mu_g$, and thermal conductivity, $k_g$, of mixtures of gasses can be described by the correlations provided by Wilke's formula \citep{Wilke1950} and the Mason-Saxena modified Wassiljewa's equation \citep{Poling2001Book}:
\begin{subequations}
\begin{align}
    \mu_g &= \mu_g(T,C) = \sum_i\frac{x_i\mu_{g,i}(T)}{\sum_jx_j\phi_{ij}(T)},\quad x_i = \frac{C_i}{\sum_jC_j} \\
    k_g &= k_g(T,C) = \sum_i\frac{x_ik_{g,i}}{\sum_jx_j\phi_{ij}(T)},\\
    \phi_{ij}(T) &= \bigg(1+\sqrt{\frac{\mu_{g,i}(T)}{\mu_{g,j}(T)}}\sqrt[4]{\frac{M_j}{M_i}}\bigg)^2\bigg(2\sqrt{2}\sqrt{1+\frac{M_i}{M_j}}\bigg)^{-1}.
\end{align}
\end{subequations}
$x_i$ is the mole fraction of gas compound $i$.
The temperature-dependent viscosity of a pure gas, $\mu_{g,i}$, is given by the correlation \citep{Sutherland1893}
\begin{align}
    \mu_{g,i}(T) = \mu_{0,i} \left(\frac{T}{T_0}\right)^{\frac{3}{2}}\frac{T_0+S_{\mu,i}}{T+S_{\mu,i}}.
\end{align}
Table \ref{tab:Data-Coeff-gas} provides the two measurement sets for calibrating $S_{\mu, i}$. 

The viscosity of the suspended solid-gas mixture, $\mu_m$, is given by the extended Einstein equation of viscosity \citep{TODA2006}:
\begin{align}
    \mu_m &= \mu_g\frac{1+\phi/2}{1-2\phi}, \quad \phi = \frac{V_s}{V_\Delta}=\hat{V}_s.  \label{eq:mue}
\end{align}

For the thermal conductivity, $k_s$, of the mixture of solid materials, we assume the mixing is equivalent to layers of material with length $\Delta z_i$ and constant cross area.
Thus, the thermal conductivity is given by the serial thermal conductivity of layered materials \citep{Perry}:
\begin{align}
    \frac{1}{k_s} = \sum_i\frac{V_{s,i}}{V_s}\frac{1}{k_{s,i}}, \quad \frac{V_{s,i}}{V_{s}} \approx \frac{\Delta z_i}{\Delta z}.
\end{align}
Each layer thickness ratio is approximated by its volumetric ratio of the solid volume.
As the air blows through the bulk in the cooler, we add the gas conductivity to the solid conductivity, scaled by the porosity:
\begin{align}
    \frac{1}{k_s} = \phi_p\frac{1}{k_{g}} + (1-\phi_p)\sum_i\frac{V_{s,i}}{V_s}\frac{1}{k_{s,i}},
\end{align}
where $V_s$ includes the porosity through the bulk density.

The thermal conductivity of the solid-gas mixture, $k_m$, is given by the same assumption with the addition of a gas layer,
\begin{align}
    \frac{1}{k_m} = \frac{V_{g}}{V_\Delta}\frac{1}{k_g}+ \sum_i\frac{V_{s,i}}{V_\Delta}\frac{1}{k_{s,i}}.
\end{align}

\subsection{Geometry}
Each module in the process relies on its specific geometry and parameters.

\subsubsection{Cyclones}
\begin{figure}[tb]
    \centering
    \begin{subfigure}[b]{0.23\textwidth}
    \centering
        \begin{tikzpicture}        
            \draw[pattern=north east lines, pattern color=brown!50] (0,-0.8) -- (0,2) arc (180:-120:2 cm) -- (1,-0.8) ; 
            \draw[pattern=north east lines, pattern color=orange!50] (0.1,-0.8) -- (0.1,2) arc (180:-125:1.9 cm) -- (0.9,-0.8) ;
            \draw[thick,-] (0.2,-0.8) -- (0.2,2) arc (180:-130:1.8 cm) -- (0.8,-0.8) ;
            \fill[white] (0.2,-0.8) -- (0.2,2) arc (180:-130:1.8 cm) -- (0.8,-0.8) ;
            
            \fill[blue!40!white] (2,2) circle (0.1 cm);           
    
            \draw[<->] (2,2) -- node[anchor=south]{$r_c$} (3.28,3.28) ;
            \draw[<->] (2,2) -- node[anchor=east]{$r_r$} (2.,3.9) ;
            \draw[<->] (2,2) -- node[anchor=south]{$r_w$} (4.0,2.0) ;

            \draw[<->] (0.5,-0.9)--node[anchor=south,scale=0.5]{$r_{in}$}(2,-0.9);
            \draw[<->] (0.2,-0.7)--node[anchor=south,scale=0.5]{$w_{in}$}(0.8,-0.7);
            \draw[<->] (-0.1,-0.8)--node[anchor=east,scale=0.5]{$l_{in}$}(-0.1,0.7);
        \end{tikzpicture}    
        \caption{Cross section.}
        \label{fig:cross_cyclone}
    \end{subfigure} 
    \begin{subfigure}[b]{0.23\textwidth}
    \centering
    \resizebox{1\textwidth}{!}{%
    \begin{tikzpicture} 

        \draw[pattern=north east lines, pattern color=brown!50] (0.6,4.0) -- (1.4,4) -- (1.4,4.1) -- (1.5,4.1) -- (1.5,3.6) -- (1.4,3.6) -- (1.4,3.9)-- (0.6,3.9) -- cycle;
        \draw[pattern=north east lines, pattern color=orange!50] (0.6,3.9) -- (1.4,3.9) --  (1.4,3.8)-- (0.6,3.8) -- cycle;
        \draw[pattern=north east lines, pattern color=brown!50] (1,3.3) -- (0.6,3.3) -- (0.6,3.4) -- (1,3.4) --  (1.1,3.4) -- (1.1,2) -- (1.6,1.1) -- (1.6,1) -- (1,2) -- cycle;
        \draw[pattern=north east lines, pattern color=orange!50] (1.1,3.4) -- (0.6,3.4) -- (0.6,3.5) -- (1,3.5) --  (1.2,3.5) -- (1.2,2) -- (1.6,1.2) -- (1.6,1.1) -- (1.1,2) -- cycle;
        \draw[pattern=north east lines, pattern color=brown!50] (2.5,4.0) -- (2.1,4) -- (2.1,4.1) -- (2.0,4.1) -- (2.0,3.6) -- (2.1,3.6) -- (2.1,3.9) -- (2.4,3.9) -- (2.4,2) -- (1.9,1.1) -- (1.9,1) -- (2.5,2) -- cycle; 
        \draw[pattern=north east lines, pattern color=orange!50] (2.4,3.9) -- (2.1,3.9) -- (2.1,3.8) -- (2.3,3.8) -- (2.3,2) -- (1.9,1.2) -- (1.9,1.1) -- (2.4,2) -- cycle; 

        \draw[->] (0.5,3.65)node[anchor=east,scale=0.5]{Inflow} -- (0.7,3.65);
        \draw[->] (1.75,1.1) -- (1.75,0.8)node[anchor=north,scale=0.5]{Separated meal outflow};
        \draw[->] (1.75,4.0) -- (1.75,4.2)node[anchor=south,scale=0.5]{Gas-Meal outflow};

        \draw[<->] (2.8,3.8)--node[anchor=west,scale=0.5]{$h_t$}(2.8,1.2);
        \draw[<->] (2.9,2)--node[anchor=west,scale=0.5]{$h_c$}(2.9,1.2);
        \draw[<->] (1.6,1.2)--node[anchor=south,scale=0.5]{$d_d$}(1.9,1.2);
        \draw[<->] (1.5,3.6)--node[anchor=south,scale=0.5]{$d_x$}(2,3.6);
        \draw[<->] (0.9,3.5)--node[anchor=west,scale=0.5]{$h_{in}$}(0.9,3.8);
        \draw[<->] (2.9,3.6)--node[anchor=west,scale=0.5]{$h_{x}$}(2.9,3.8);

        \draw[thick,-] (1.05,3) -- node[anchor=north east,scale=0.5]{Wall} (0.8,3.0) ;
        \draw[thick,-] (1.15,2.5) -- node[anchor=north east,scale =0.5]{Refractory} (0.8,2.5) ;

        \draw[<-,color=red,decoration={aspect=0.61, segment length=3mm, amplitude=0.4cm,coil},decorate] (1.75,1.4) --(1.75,3.6);
        \draw[color=blue,decoration={aspect=0.61, segment length=3mm, amplitude=0.2cm,coil},decorate,arrows = {<[bend]-}] (1.75,3.5) --(1.75,1.5);        
    \end{tikzpicture}
    }
    \caption{Axial profile.}
    \label{fig:axial_cyclone}
    \end{subfigure}
    \caption{Geometric profiles of the cyclone: 
    (a) the horizontal cross section at entrance level,
    (b) the vertical axial profile.} \label{fig:profiles_cyclone}
\end{figure}
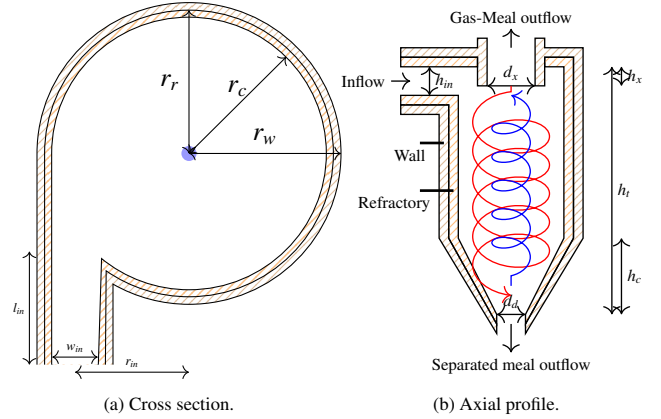

The cyclone chamber is considered as one collective volume.
Fig. \ref{fig:profiles_cyclone} illustrates the geometry of the cyclone.
The volume,\ $V_\Delta$, and internal surface area, $A_{cy}$, are given by
\begin{subequations}
\begin{align}
    V_{\Delta} =\pi(r_c^2(h_t-h_c)& + \frac{h_c}{3}(r_c^2+r_x^2+r_cr_x)- r_x^2h_{x}),\\
    \begin{split}
    A_{cy} = 2 \pi r_c(h_t-h_c)& + \pi(r_c^2-r_x^2) \\
    &  + \pi (r_c+r_d)\sqrt{(r_c-r_d)^2+h_c^2}.     
    \end{split}    
\end{align}
\end{subequations}
The collection area for the separation of the suspended solids from the gas is
\begin{small}
\begin{subequations}
    \begin{align}
        A_{sep} & = 2 \pi r_c(h_t-h_c) + \pi (r_c+r_2)\sqrt{(r_c-r_2)^2+h_{c,1}^2},\\
        r_2 &= r_c - \frac{h_{c,1}}{h_{c}}(r_c-r_d), \quad h_{c,1}  = h_c/2.
    \end{align}
    \end{subequations}
\end{small}

\subsubsection{Calciners and risers}
\begin{figure}[tb]
    \centering
    \begin{subfigure}[b]{0.24\textwidth}
    \centering
        \begin{tikzpicture}        
            \draw[pattern=north east lines, pattern color=brown!50] (2,2) circle (2 cm);
            \draw[pattern=north east lines, pattern color=orange!50] (2,2) circle (1.8 cm);
            \draw[thick,-] (2,2) circle (1.5 cm);
            \fill[white] (2,2) circle (1.5 cm);
            \fill[blue!40!white] (2,2) circle (0.1 cm);
    
            \draw[thick,<->] (2,2) -- node[anchor=south]{$r_c$} (3.05,3.05) ;
            \draw[thick,<->] (2,2) -- node[anchor=south]{$r_r$} (3.8,2) ;
            \draw[thick,<->] (2,2) -- node[anchor=south]{$r_w$} (0.6,0.6) ;

            \draw[-] (3.8,2.5) -- (4,3.0) node[anchor=south]{$A_w$};
            \draw[-] (3.5,1.5) -- (4.5,1.5) node[anchor=south]{$A_r$};
            \draw[-] (2.0,3.0) -- (3.5,4.0) node[anchor=west]{$A_c$};
        \end{tikzpicture}    
        \caption{Cross section}
        \label{fig:cross_calciner}
    \end{subfigure}%
    \begin{subfigure}[b]{0.24\textwidth}
    \centering
    \resizebox{1\textwidth}{!}{%
    \begin{tikzpicture}   
        \draw[pattern=north east lines, pattern color=brown!50] (0,1.0) -- (0,3) -- (0.5,4) -- (0.6,4) -- (0.1,3) -- (0.1,1) -- (0.6,0) -- (0.5,0)-- cycle;
        \draw[pattern=north east lines, pattern color=orange!50] (0.1,1.0) -- (0.1,3) -- (0.6,4) -- (0.7,4) -- (0.2,3) -- (0.2,1) -- (0.7,0) -- (0.6,0)--cycle;
        
        \draw[pattern=north east lines, pattern color=brown!50] (2.2,1.0) -- (2.2,3) -- (1.7,4) -- (1.8,4) -- (2.3,3) -- (2.3,1) -- (1.8,0) -- (1.7,0)-- cycle;
        \draw[pattern=north east lines, pattern color=orange!50] (2.1,1.0) -- (2.1,3) -- (1.6,4) -- (1.7,4) -- (2.2,3) -- (2.2,1) -- (1.7,0) -- (1.6,0)--cycle;

        \draw[thick,-] (2.25,3) -- node[anchor=south west,scale=0.5]{Wall} (2.5,3.0) ;
        \draw[thick,-] (2.15,2.5) -- node[anchor=south west,scale =0.5]{Refractory} (2.5,2.5) ;

        \draw[thick,->] (-0.5,-0.1) -- (-0.5,4.5)node[anchor= west,scale =0.5]{y};
        \draw[thick,-] (-0.6,-0.0) -- (-0.4,0)node[anchor= west,scale =0.5]{0};
        \draw[thick,-] (-0.6,4) -- (-0.4,4)node[anchor= west,scale =0.5]{$h_{tot}$};
        \draw[thick,-] (-0.6,3) -- (-0.4,3)node[anchor= west,scale =0.5]{$h_{cu}$};
        \draw[thick,-] (-0.6,1) -- (-0.4,1)node[anchor= west,scale =0.5]{$h_{cl}$};

        \draw[-] (-0.6,2.4)node[anchor= east,scale =0.5]{$y_{k+\frac{1}{2}}$} -- (2.3,2.4);
        \draw[-] (-0.6,2.0)node[anchor= east,scale =0.5]{$y_{k-\frac{1}{2}}$} -- (2.3,2.0);

        \draw[->] (1.25,1.9) --  (1.25,2.2);
        \draw[->] (1.25,2.3) --  (1.25,2.6);

        \draw[<->] (1.22,0) -- node[anchor=north,scale =0.5]{$r_l$} (1.6,0);
        \draw[<->] (1.22,4) -- node[anchor=north,scale =0.5]{$r_u$} (1.6,4);
        
    \end{tikzpicture}
    }
    \caption{Axial profile, with segment \\notation for the $k$-th volume.}
    \label{fig:axial_calciner}
    \end{subfigure}
    \caption{Diagrams of the calciner profiles. The diagrams illustrate the dimensions and flow direction.}\label{fig:profiles_calciner}
\end{figure}
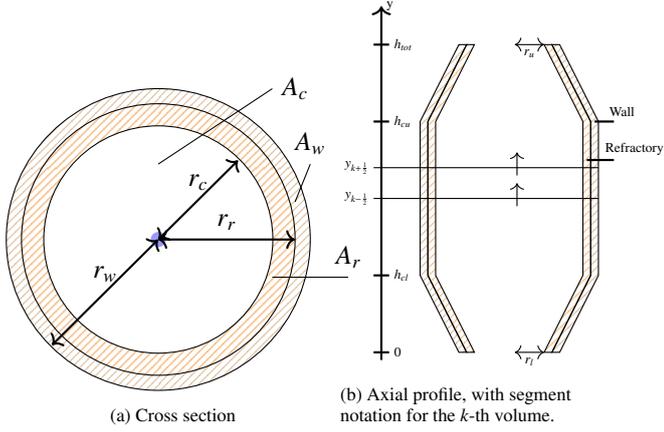

The calciner and the risers are considered to be straight pipes with varying diameters.
The calciner is a cylinder with truncated cones at each end. 
The risers are cylinders with angled pipes on top. 
The angled pipes are modeled as truncated cones, where the height of the cones equals the pipe length and the top areas equal the cross-sectional area of the pipes' end.
Fig. \ref{fig:profiles_calciner} illustrates the geometry of a calciner.
The segment volumes of the calciner and risers are thus given by
\begin{align}
    V_{\Delta}(k) = \pi r_c^2h_c(k) &+ \frac{\pi}{3}(r_c^2+r_u^2(k) +r_cr_u(k))h_u(k)\nonumber\\ &+ \frac{\pi}{3}(r_c^2+r_l^2(k) +r_cr_l(k))h_l(k).
\end{align}
The formula for the height of the cylinder and cone parts of the segment are 
\begin{subequations}
\begin{align}
    h_c(k) &= (y_{k+\frac{1}{2}}-h_u(k))-(y_{k-\frac{1}{2}}+h_l(k)),\\
    h_u(k) &= \max(y_{k+\frac{1}{2}}-h_{cu},0),\\
    h_l(k) &= \max(h_{cl}-y_{k-\frac{1}{2}},0),
\end{align}
\end{subequations}
where $h_{cl}$ is zero for the risers.
The smaller cone radius of the segment is given by
\begin{subequations}
\begin{align}
    r_u(k) &= r_u+\frac{r_c-r_u}{h_{cu}}h_u(k),\\
    r_l(k) &= r_l+\frac{r_c-r_l}{h_{cl}}h_l(k).
\end{align}
\end{subequations}
The segment volumes for the refractory and walls are computed by the relations
\begin{subequations}
\begin{align}
    V_{r,\Delta}(k) &= V_{\Delta}(k)|_{r_c=r_r} - V_{\Delta}(k),\\
    V_{w,\Delta}(k) &= V_{\Delta}(k)|_{r_c=r_w} - V_{\Delta}(k)|_{r_c=r_r}.
\end{align}
\end{subequations}
The internal surface area of each segment, $A_c$, is given by
\begin{equation}
\begin{split}
    A_c(k) = 2 \pi r_c h_c(k) &+ \pi (r_c + r_u(k)) \sqrt{h_u^2(k) + (r_c-r_u(k))^2}\\
    & + \pi (r_c + r_l(k)) \sqrt{h_l^2(k) + (r_c-r_l(k))^2}.
\end{split}
\end{equation}

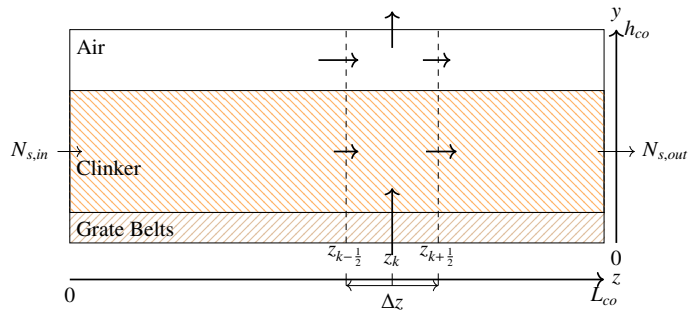
\begin{figure}
    \centering
     \resizebox{0.5\textwidth}{!}{%
     \begin{tikzpicture}   
        \draw[thick,->] (0,-0.6)node[anchor=north]{$0$} -- (8.7,-0.6)node[anchor=north]{$L_{co}$}node[anchor=west]{$z$};
        \draw[thick,->] (8.9,0.0)node[anchor=north]{$0$} -- (8.9,3.5)node[anchor=west]{$h_{co}$}node[anchor=south]{$y$};
        \draw[pattern=north east lines, pattern color=brown!50] (0,0.0) rectangle (8.7,0.5);
        \fill[white] (0,0.5) rectangle (8.7,3.5);
        \draw (0,0.5) rectangle (8.7,3.5);        

        \draw[pattern=north west lines, pattern color=orange!60]  (0,2.5) -- (0,0.5) -- (8.7,0.5) -- (8.7,2.5)-- cycle;

        \draw[->]  (-0.2,1.5) -- node[left=6pt]{$N_{s,in}$} (0.2,1.5);
        \draw[->]  (8.6,1.5) -- (9.2,1.5)node[right]{$N_{s,out}$};
        
        \draw (5.25,-0.7) -- (5.25,-0.5) node[anchor=south]{$z_{k}$};
        \draw[dashed] (4.5,0) -- (4.5,3.5);
        \draw (4.5,-0.7) -- (4.5,-0.5) node[anchor=south]{$z_{k-\frac{1}{2}}$};
        \draw[dashed] (6.0,0) -- (6.0,3.5);
        \draw (6.0,-0.7) -- (6.0,-0.5)node[anchor=south]{$z_{k+\frac{1}{2}}$};
        \draw[<->] (4.5,-0.7) -- node[anchor=north]{$\Delta z$} (6.0,-0.7);

        \draw[thick,->] (5.25,3.2) -- (5.25,3.8);
        \draw[thick,->] (4.05,3.0) -- (4.7,3.0);
        \draw[thick,->] (5.75,3.0) -- (6.2,3.0);
        \draw[thick,->] (5.25,-0.2) -- (5.25,0.9) ;
        
        \draw[thick,->] (4.3,1.5) --  (4.7,1.5);
         \draw[thick,->] (5.8,1.5) -- (6.3,1.5);

        \node[anchor=south west,inner sep=0] at (0.1,0.1){Grate Belts};
        \node[anchor=south west,inner sep=0] at (0.1,1.1){Clinker};
        \node[anchor=south west,inner sep=0] at (0.1,3.1){Air};
    \end{tikzpicture}
     }
\caption{Diagram of the cooler axial profile. The arrows show the cross-current flows of clinker and air.}
    \label{fig:coolergeo1}
\end{figure}

\subsubsection{Cooler} 
The cooler is considered to have a rectangular shape with width $w_{co}$, height $h_{co}$, and length $L_{co}$.
Defining $n$ segments along the length, each segment has length $\Delta z=L_{co}/n$.
The segment volume is thus given by $V_\Delta(k)=w_{co} h_{co} \Delta z$.
The end placements of the grates in the cooler are given as $p_{co}$, a vector with percentages of the cooler length.
Fig. \ref{fig:coolergeo1} illustrates the axial geometry of the cooler.

\subsubsection{Kiln}
\begin{figure}
    \centering
    \begin{subfigure}[b]{0.5\textwidth}
    \centering
        \begin{tikzpicture}        
            \draw[pattern=north east lines, pattern color=brown!50] (2,2) circle (2 cm);
            \draw[thick,-] (2,2) circle (1.5 cm);
            \fill[white] (2,2) circle (1.5 cm);
            \fill[blue!40!white] (2,2) circle (0.1 cm);
            \draw[pattern=north west lines, pattern color=orange!60]  (1.5,0.6) arc (-110:0:1.5cm)  -- cycle;
    
            \draw[ultra thick, ->] (2,4.2)  arc (90:150:2.2cm) node[anchor=south east]{$\omega$};\draw[] (2,4.4);
            \draw[thick,<->] (2,2) -- node[anchor=south]{$r_c$} (3.05,3.05) ;
    
            \draw[thick,<->] (1.5,0.6) -- node[anchor=south]{$L_c$} (3.5,2) ;        
            \draw[-] (1.5,0.6) -- (2,2) ;
            \draw[-] (3.5,2) --  (2,2) ;
            \draw (2.2,2)node[anchor=north]{$\theta$} arc (0:-112:0.2cm) ;
            \draw[thick,<->] (2.5,1.3)  --  node[anchor=south west]{$h$} (2.88,0.775) ;
    
            \draw[dashed,-] (1.5,0.6) -- (0.65,0) ;
            \draw[dashed,-] (2,0) -- (0.65,0) ;
            \draw (1.0,0)node[anchor=north west]{$\xi$} arc (0:35:0.35cm) ;
    
            \draw[-] (3.7,2.5) -- (4,3.0) node[anchor=south]{$A_r$};
            \draw[-] (3.2,1.5) -- (4.5,1.5) node[anchor=south]{$A_s$};
            \draw[-] (2.0,3.0) -- (3.5,4.0) node[anchor=west]{$A_g$};
        \end{tikzpicture}    
        \caption{Cross section.}
        \label{fig:crossKiln}
    \end{subfigure}
    
    \begin{subfigure}[b]{0.5\textwidth}
    \centering
    \resizebox{1\textwidth}{!}{%
    \begin{tikzpicture}   
        \draw[thick,->] (0,-0.6)node[anchor=north]{$0$} -- (8.7,-0.6)node[anchor=north]{$L$}node[anchor=west]{$z$};
        \draw[pattern=north east lines, pattern color=brown!50] (0,0.0) rectangle (8.7,4.0);
        \fill[white] (0,0.5) rectangle (8.7,3.5);
        \draw (0,0.5) rectangle (8.7,3.5);

        \draw[pattern=north west lines, pattern color=orange!60]  (0,1.9) -- (8.7,1.2) -- (8.7,0.5) -- (0,0.5)-- cycle;
        \draw (8.95,0.0625) -- (6.7,-0.5); 
        \draw  (7.3,-0.3) arc (-112:-180:0.3cm) node[anchor= north east]{$\psi$};

        \draw[dashed,blue!40!white] (0,2) -- (9,2);
        \draw[dashed,blue!40!white] (6.9,1.4) -- (8.3,1.4);
        \draw (8.2,1.4)node[anchor=west]{$\phi$} arc (0:-35:0.25cm) ;

        \draw (2.5,-0) rectangle node[anchor=south]{$z_{k}$}(4.0,4);
        \draw (2.5,-0.7) -- (2.5,-0.5) node[anchor=south]{$z_{k-\frac{1}{2}}$};
        \draw (4.0,0) rectangle node[anchor=south]{$z_{k+1}$} (5.5,4) ;
        \draw (4.0,-0.7) -- (4.0,-0.5) node[anchor=south]{$z_{k+\frac{1}{2}}$};
        \draw (5.5,-0.7) -- (5.5,-0.5) node[anchor=south]{$z_{k+\frac{3}{2}}$};
        \draw[<->] (2.5,-0.7) -- node[anchor=north]{$\Delta z$} (4.0,-0.7);
        \draw[<->] (4.0,-0.7) -- node[anchor=north]{$\Delta z$} (5.5,-0.7);
        
        \draw[thick,<->] (4.0,0.5)  --node[anchor=west]{$h_{k+\frac{1}{2}}$}  (4.0,1.6) ;
        
        \draw[thick,<-] (2.25,2.8) -- (2.75,2.8);
        \draw[thick,<-] (5.25,2.8) -- (5.75,2.8);
        \draw[thick,->] (2.25,1.1) -- (2.75,1.1);
        \draw[thick,->] (5.25,1.0) -- (5.75,1.0);
        \draw[thick,<->] (3.2,1.4) -- (3.2,1.8);

        \node[anchor=south west,inner sep=0] at (0.1,0.1){Refractory};
        \node[anchor=south west,inner sep=0] at (0.1,3.6){Refractory};
        \node[anchor=south west,inner sep=0] at (0.1,1.1){Solid};
        \node[anchor=south west,inner sep=0] at (0.1,3.1){Gas};
    \end{tikzpicture}
    }
    \caption{Kiln axial profile.}
    \label{fig:axial}
    \end{subfigure}
    \caption{Diagram of kiln profiles. The arrows show 
    the flow directions.}\label{fig:profilesKiln}
\end{figure}
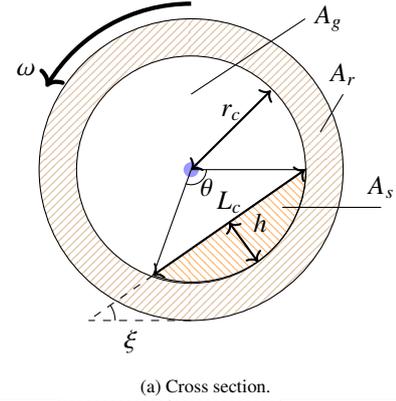
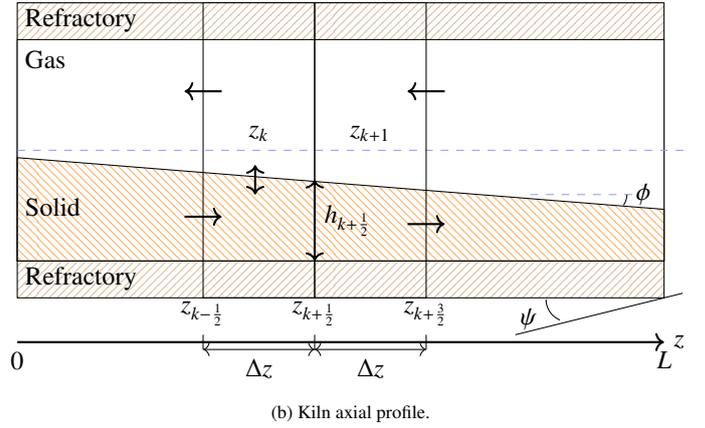

The kiln module differs from the other modules since the solid volume across the kiln defines the geometric aspect of each segment. Fig. \ref{fig:profilesKiln} illustrates the kiln geometry.
The cross area of the solid phase is given by the solid volume
\begin{align}
    V_s(z_k) = \int_{z_{k-\frac{1}{2}}}^{z_{k+\frac{1}{2}}} A_s(z) dz \approx A_s(z_k) \Delta z.
\end{align}
From the cross area, $A_s$, and assuming the solids form a circle segment, the fill angle, $\theta$, is defined by the geometry \citep{CircSeg}
\begin{align}
    A_s(z) &= \frac{r_c^2}{2}(\theta(z) - \text{sin}(\theta(z)) )\label{eq:geoAlg}.
\end{align}
The surface width, $L_c$, the bed height, $h$, and the bed slope angle, $\phi$, are then given by
\begin{subequations}
\begin{align} 
    L_c(z) &= 2 r_c\text{sin}\bigg(\frac{\theta(z)}{2}\bigg),\quad h(z) = r_c\bigg(1-\text{cos}\bigg(\frac{\theta(z)}{2}\bigg)\bigg),\\          
    \phi(z) &= \text{atan}\bigg(-\pdv{h(z)}{z}\bigg).
\end{align}
\end{subequations}
The gas cross-sectional area, $A_g$, and the surface areas between each phase are given by 
\begin{subequations}
\begin{align}
    A_g(z) & = \pi r_c^2 - A_s(z),\\
    A_{gs}(z_k) &= \int_{z_{k-\frac{1}{2}}}^{z_{k+\frac{1}{2}}} L_c(z) dz,\\
    A_{rs}(z_k) &= \int_{z_{k-\frac{1}{2}}}^{z_{k+\frac{1}{2}}} r_c\theta(z) dz,\\
    A_{gr}(z_k) &= 2\pi r_c\Delta z - A_{rs}(z_k),
\end{align}
\end{subequations}
where $A_{gs}$ is the gas-solid surface area, $A_{gr}$ is the gas-refractory surface area, and $A_{rs}$ is the refractory-solid surface area.

\subsection{Stoichiometry and kinetics}\label{secsub:kinetic}
The chemical reactions in the pyro-process are included in the model through the production rate vector, $R$. The production rates, $R$, are computed by the reaction rate vector, $r = r(T,P,C)$, and the stoichiometric matrix, $\nu$:
\begin{align}
    R = \begin{bmatrix}
        R_s\\R_g
    \end{bmatrix}= \nu^Tr.
\end{align}
 $R_s$ is the production rate vector of the solid phase (\ce{CaCO3}, \ce{CaO}, \ce{SiO2}, \ce{AlO2}, \ce{FeO2}, \ce{C_2S}, \ce{C_3S}, \ce{C_3A}, and \ce{C_4AF}) and $R_g$ is the production rate vector of the gas phase (\ce{CO_2}, \ce{N_2}, \ce{O_2}, \ce{Ar}, \ce{CO}, \ce{H_2}, \ce{H_2O}), and $C_{sus}$ (suspended carbon).
The stoichiometry matrix, $\nu$, describes the relations of the included reactions.

The included clinker formation reactions are
\begin{subequations}
\begin{alignat}{5}
    \text{$\#1$: }& & \ce{CaCO3} &\rightarrow \ce{CO2} + \ce{CaO}, \quad & r_1,\label{eq:r1}\\
    \text{$\#2$: }& & 2\ce{CaO} + \ce{SiO_2} &\rightarrow \ce{C_2S}, \ & r_2,\\
    \text{$\#3$: }& &\ce{CaO} + \ce{C_2S}&\rightarrow \ce{C_3S}, \ & r_3,\\
    \text{$\#4$: }& &3\ce{CaO} + \ce{Al_2O_3}&\rightarrow \ce{C_3A}, \ & r_4,\\
    \text{$\#5$: }& &4\ce{CaO} + \ce{Al_2O_3} + \ce{Fe_2O_3}&\rightarrow \ce{C_4AF}, \ & r_5,\\
    \text{$\#6$: }& &\ce{C_3S}&\rightarrow \ce{C_2S} + \ce{CaO}, \ & r_6,
    \end{alignat}
\end{subequations}    
while the included reactions for fuel combustion are  
\begin{subequations}
    \begin{alignat}{6}
   \text{$\#$}&\text{$7$: }& &2\ce{C} +\ce{ O_2} &&\rightarrow 2\ce{CO}, \quad & r_7,\\
   \text{$\#$}&\text{$8$: }& &\ce{C} + \ce{H_2O}&&\rightarrow \ce{CO} + \ce{H_2}, \quad & r_{8}, \\
   \text{$\#$}&\text{$9$: }& &\ce{ C} + \ce{CO_2}&&\rightarrow 2\ce{CO}, \quad & r_{9},\\
   \text{$\#$}&\text{$10$: }\ & & 2\ce{CO}+\ce{O_2}&&\rightarrow 2\ce{CO_2}, \quad & r_{10}, \\
   \text{$\#$}&\text{$11$: }\ & & \ce{CO} + \ce{H_2O} &&\rightarrow \ce{CO_2} + \ce{H_2}, \quad & r_{11},\\
   \text{$\#$}&\text{$12$: }\ & & 2\ce{H_2}+\ce{O_2}&&\rightarrow 2\ce{H_2O}, \quad & r_{12}.
\end{alignat}
\end{subequations}
The reaction rate functions, $r_j(T, P, C)$, used in this paper are given by expressions of the form
\begin{align}
    r = k(T)\prod_lP_l^{\beta_l}C_l^{\alpha_l},\quad k(T) = k_{r}T^ne^{-\frac{E_{A}}{RT}}.
\end{align}
 $k(T)$ is the modified Arrhenius expression; $C_l$ is the concentration expressed on the volume basis mol/L; $\alpha_l$ is either the stoichiometric-related or experimental-based power coefficient; $\beta_l$ is the power of the partial pressure $P_l = (C_l/\sum_j C_j ) P$. 
 
  Table \ref{tab:allreactions} shows the reaction coefficients found in the literature for the clinker and fuel reactions.
  Table \ref{tab:spang} shows alternative coefficients by \cite{Spang} for reaction $r_1-r_5$.
  The order of magnitude difference for $k_r$ in $r_1$, $10^{17}$, illustrates the empirical nature of reaction rate coefficients. Thus the coefficients in the literature should be considered guiding values for tuning the model to the system of interest.
\begin{table}[t]
\centering
\begin{scriptsize}
    \caption{Reaction rate coefficients.}
    \begin{tabular}{c c|c c c | c c c | c}\hline
         & & $k_r$ &$n$&$E_{A}$ & $\alpha_1$ & $\alpha_2$ & $\alpha_3$ & $\beta_2$\\ \hline
    Reactions & Units & \small{$\frac{[r_i]}{[C]^{\Sigma\alpha}[P]^{\Sigma\beta}}$} & 1& $\frac{\text{kJ}}{\text{mol}}$& 1& 1& 1& 1\\ \hline 
    & & & & & & & & \\[-1em]
    $r_1$ $^a$& $\frac{\text{kg}}{\text{m}^3\text{s}}$& $10^{8}$& & $175.70$& 1& &\\  & & & & & & & &\\[-1em]
    $r_2$ $^a$& $\frac{\text{kg}}{\text{m}^3\text{s}}$&$10^{7}$ & & $240.00$& 2& 1&\\  & & & & & & & &\\[-1em]
    $r_3$ $^a$& $\frac{\text{kg}}{\text{m}^3\text{s}}$& $10^{9}$& & $420.00$& 1& 1&\\  & & & & & & & &\\[-1em]
    $r_4$ $^a$& $\frac{\text{kg}}{\text{m}^3\text{s}}$&$10^{8}$ & & $310.00$& 3& 1&\\  & & & & & & & &\\[-1em]
    $r_5$ $^a$& $\frac{\text{kg}}{\text{m}^3\text{s}}$& $10^{8}$& & $330.00$& 4& 1&1\\   & & & & & & & &\\[-1em]
    $r_6$ $^b$& s$^{-1}$& $0.09$ & & $96.58$ & 1& & \\  & & & & & & & &\\[-1em]
    \hline
    $r_{7}$ $^c$& $\frac{\text{mol}}{\text{m}^3s}$ & $8.82\cdot10^{13}$ &  &239.00 & $1$  & $0.5$  & &  \\ & & & & & & & &\\[-1em]
    $r_{8}$ $^d$& $\frac{\text{mol}}{\text{m}^3s}$& $2.62\cdot10^8$ &  & 237.00 & 1 &  & &  0.57 \\  & & & & & & & &\\[-1em]
    $r_{9}$ $^d$& $\frac{\text{mol}}{\text{m}^3s}$& $3.1\cdot10^6$ &  & 215.00 & 1 &   & & 0.38 \\ 
    $r_{10}$ $^e$& $\frac{\text{mol}}{\text{m}^3s}$&$7.0\cdot10^4$ &  & $66.51$& $1$ & $1$ & & \\  & & & & & & & &\\[-1em]
    $r_{11}$ $^f$& $\frac{\text{mol}}{\text{m}^3s}$& $2.75\cdot10^6$ &  & 83.68 & 1& 1& &  \\  & & & & & & & &\\[-1em]
    $r_{12}$ $^g$& $\frac{\text{mol}}{\text{m}^3s}$& $1.37\cdot10^{6}$ & 0.51 & $295.48$ & 1 & 1 & &  \\  
     \hline
    \end{tabular} 
    \label{tab:allreactions}
\end{scriptsize}
    \footnotesize{ All $\beta_1$ and $\beta_3$ are zero. $^a$ \cite{Mastorakos1999CFDPF}. $^b$ \cite{CS3toCS2}, estimated using least-square and the Jander-data. $^c$ \cite{Walker1985}. $^d$ \cite{BASU2018211}. $^e$ \cite{Guo2003}. $^f$ \cite{JONES1988}. $^g$ \cite{Karkach1999}. }
       
\end{table}

\begin{table}[t]
\centering
    \caption{Reaction rate coefficients of \cite{Spang} for clinker formulation.}
    \begin{footnotesize}  
    \begin{tabular}{c c | c | c | c c c }\hline
    && $k_r$ &  $E_{A}$ & $\alpha_1$ & $\alpha_2$ & $\alpha_3$ \\
    \hline
    Reactions &Units  & h$^{-1}\cdot [C]^{-\Sigma\alpha}$ & $\frac{\text{kJ}}{\text{mol}}$ & 1 & 1 & 1 \\
    \hline & & & & & &\\[-1em]
    $r_1$ & h$^{-1}$& $1.64\cdot 10^{35}$  & $804.8$&  1& & \\
    $r_2$ & h$^{-1}$& $14.8\cdot 10^{8}$ & $193.1$&  2& 1& \\
    $r_3$ & h$^{-1}$& $4.8\cdot 10^{8}$& $255.9$&  1& 1& \\
    $r_4$ & h$^{-1}$& $300\cdot 10^{8}$ & $193.8$&  3& 1& \\
    $r_5$ & h$^{-1}$& $30\cdot10^{11}$& $184.9$ &  4& 1&1 \\
    \hline
    \end{tabular}  
    \end{footnotesize}
    
    \footnotesize{All $\beta$ and $n$ are zero.
    The units of the reactions are $[\frac{1}{\text{h}}]$ and $[\frac{\text{kJ}}{\text{mol}}]$ for the activation energy $E_{A}$.}
       \label{tab:spang}
\end{table}

\subsection{Heat transfer}
The transfer of heat between each phase is given by heat convection, thermal radiation, and phase transitions.
As the kiln and cooler modules have separate internal energy phases for the solid and gas phases, the calcination process releases \ce{CO2} from the solid phase to the gas phase, thus transferring heat. The \ce{CO2} transfer is given by \eqref{eq:r1} with the corresponding heat transfer by
\begin{align}
    J_{sg} &= H(T_s,P,R_{g,1}), \quad  R_{g,1} = \nu_{g,1}^Tr_1.
\end{align}
$R_{g,1}$ is the production rate of gasses in reaction 1, and $\nu_{g,1}$ is the stoichiometric vector of the gasses in reaction 1.
\subsubsection{Heat convection}
Heat transfer due to convection between two phases $i$ and $j$ are given by
\begin{align}
    Q_{ij}^{cv} &= A_{ij}\beta_{ij}(T_i-T_j).
\end{align}
The general total specific heat transfer, $A\beta$, is given by  \citep{CengelBook2002}
\begin{align}
    A\beta = \left(\frac{1}{A_0\beta_0} + \sum_{i=1}^{n-1}\frac{dx_i}{k_iA_i}+ \frac{1}{A_n\beta_n}\right)^{-1},
\end{align}
where $A_0\beta_0$ and $A_n\beta_n$ are the convection heat transfer of fluids, and the $\sum$-term is the insulation by barriers. $dx$ and $k$ are the thickness and thermal conductivity of the barrier. $A$ is the contact area between phases.
The convective heat transfer coefficient, $\beta$, of the fluids (gas and/or solid flow) is given by 
\begin{align}
    \beta = \frac{k}{d}Nu.
\end{align}
$Nu$ is the Nusselt number, and d is the characteristic length.

For the cyclones, calciner, and risers the temperature of each phase is assumed to be located in the center of each phase. Thus, the overall heat transfer between phases is given by
\begin{subequations}
\begin{align}
    A_{mr}\beta_{mr} &= \left(\frac{1}{A_{c}\beta_m} + \frac{dx_{r_c,0.5(r_c+r_r)}}{k_r\frac{A_{c}+A_r}{2}}\right)^{-1},\\
    A_{rw}\beta_{rw} &= \left(\frac{dx_{r_r,0.5(r_c+r_r)}}{k_rA_{r}} + \frac{dx_{0.5(r_r+r_w),r_r}}{k_w\frac{A_r+A_w}{2}}\right)^{-1},\\
    A_{we}\beta_{we} &= \frac{k_wA_{w}}{dx_{r_w,0.5(r_r+r_w)}}.
\end{align}
\end{subequations}
$dx_{i,j} = \text{ln}(\frac{r_{j}}{r_i})r_i$ is the depth for curved walls with inner and outer radii $r_i$ and $r_j$. 
For the cyclones, the heat coefficient of the mixture, $\beta_m$, is given by \citep{Gupta2000}
\begin{subequations}
\begin{align}
    \beta_m &= \frac{k_m}{D_H}Nu_m,\quad D_H = \frac{4V_{\Delta}}{A_{cy}}, \quad Re = \frac{\rho_g v_{in} 2r_c}{\mu_g},\\
    \begin{split}
    Nu_m& = 702.8 + 9.5\vdot10^{-8}\frac{v_{in}}{u_{mf}}\frac{2r_c}{d_p}Re \\
    &+ (0.03 + 1.2\vdot10^{-13}\frac{v_{in}}{u_{mf}}\frac{2r_c}{d_p}Re)\frac{\rho_s}{\rho_g}\frac{c_{s}}{c_{g}}\frac{k_s}{k_g}\frac{\Delta P_c}{0.5\rho_g v_{in}^2}.
\end{split}
\end{align}
\end{subequations}
$D_H$ is the hydraulic diameter. $d_p$ is the solid particle diameter.  $u_{mf}$ is the minimum fluidization velocity (\cite{ZHOU2020Umf} suggested 0.16m/s). $Re$ is the Reynolds number. $\Delta P_c$ is the pressure drop across the cyclone.

For the calciner and the risers, the heat coefficients, $\beta_m$, are computed using the Gnielinski correlation for the Nusselt number, $Nu$, of turbulent flow in tubes \citep{Incropera}:
\begin{subequations}
\begin{align}
    \beta_m &= \frac{k_m}{D_H}Nu, \quad& Nu &= \frac{\frac{f}{8}(Re_D-1000)Pr}{1+12.7(\frac{f}{8})^\frac{1}{2}(Pr^\frac{2}{3}-1)},\\    
    D_H &= \frac{4V_{\Delta}}{A_{c}}, \quad& Re_D &= \frac{\rho_m v_c D_H}{\mu_m},\\    
    Pr &= \frac{c_{m} \mu_m}{k_m},\quad& f &=(0.79 \ln (Re_D) - 1.64)^{-2}.    
\end{align}
\end{subequations}
$Pr$ is the Prandtl number. $Re_D$ is the Reynolds number for $D_H$.

For the kiln, the refractory temperature is assumed to be located at the inner surface, i.e., no insulation part. Thus, $A_{ij}$ is simply the surface area between the phases, and the heat coefficients are given by \citep{Tscheng1979} 
\begin{subequations}
\begin{align}\label{eq:betagsKiln}
    \beta_{gs} =& \frac{k_g}{D_e}Nu_{gs}, \ Nu_{gs}=0.46Re_D^{0.535}Re_\omega^{0.104}\eta_k^{-0.341},\\
    \beta_{gr} &= \frac{k_g}{D_e}Nu_{gr}, \quad Nu_{gr}=1.54Re_D^{0.575}Re_\omega^{-0.292},\\
    \beta_{rs} &= 11.6\frac{k_s}{l_r}(\frac{\omega r_c^2\theta}{\alpha_B})^{0.3}, \ \alpha_B = \frac{k_s}{\rho_sC_{ps}},\ l_r = r_c\theta,\\
    Re_D &= \frac{\rho_gv_gD_e}{\mu_g}, \quad Re_\omega = \frac{\rho_g\omega D_e^2}{\mu_g},\\
    \eta &= \frac{\theta-\text{sin}(\theta)}{2\pi},\quad
D_e = 2r_c \frac{\pi-\frac{\theta}{2}+\frac{\text{sin}(\theta)}{2}}{\pi-\frac{\theta}{2}+\text{sin}(\frac{\theta}{2})}.
\end{align}
\end{subequations}
$D_e$ is the effective diameter. $\alpha_B$ is the thermal diffusivity. $l_r$ is the contact perimeter between the solid and refractory. $Re_D$ and $Re_\omega$ are Reynold's numbers and $\eta_k$ is the solid fill fraction.

In the cooler, the air flows through the clinker bed, thus, the surface area is the total clinker particle area, $A_p$. \cite{CUI20171297}  gives the heat transfer coefficient as
\begin{subequations}
\begin{align}
    A_p &=\frac{6}{D_p}V_s, \\
    \beta &= \frac{k_sNu}{D_p + 0.5\phi D_pNu}.
\end{align}
\end{subequations}
$\phi = 0.25$ is the clinker shape correction factor. $D_p$ is the average clinker particle diameter, typically around $40$ mm. The Nusselt Number $Nu$ is given by
\begin{subequations}
\begin{align}
    Nu &= 2+1.8Pr^{\frac{1}{3}}Re^{\frac{1}{2}},\\
    Pr &= \frac{c_{s}\mu_g}{k_g}, \quad Re = \frac{\rho_gv_yD_p}{\mu_g}.
\end{align}
\end{subequations}

\subsubsection{Thermal radiation}
Stefan-Boltzmann law describes the total energy transfer by thermal radiation from phase $i$ to phase $j$:
\begin{align}
    Q_{ij}^{rad} = A_{ij}\epsilon_i\sigma T_i^4.
\end{align}
$\sigma$ is Stefan-Boltzmann's constant. $\epsilon_i$ is the emissivity of phase $i$. Accounting for both directions, the general formulation is given by
\begin{align}
    Q_{ij}^{rad} = A_{ij}\sigma(\epsilon_i T_i^4 - \epsilon_j T_j^4).
\end{align}
This formulation is used for the calciner, risers, and cooler modules.
The thermal radiation of the cyclones is given by \citep{Gupta2000}
\begin{subequations}
\begin{align}
    Q_{mr}^{rad} &= \sigma A_{cs}F_{p-r}(T^4_c-T^4_r), \quad F_{p-r} = \frac{1}{\frac{1}{\epsilon_p}+\frac{1}{\epsilon_r}-1},\\
    Q_{we}^{rad} &= \sigma A_{w}(\epsilon_wT^4_w-\epsilon_eT^4_e), \quad A_{cs}=A_c\hat{V}_s.
\end{align}
\end{subequations}
$\epsilon_p$ is the emissivity of the solid particles.
For the kiln, the formulation of the thermal radiation is described by \citep{Mujumdar2006}
\begin{subequations}
\begin{align}
    Q_{gs}^{rad} &= \sigma A_{gs}(1+\epsilon_s)\frac{\epsilon_gT^4_g-\alpha_gT^4_s}{2},\\
    Q_{gw}^{rad} &= \sigma A_{gw}(1+\epsilon_w)\frac{\epsilon_gT^4_g-\alpha_gT^4_w}{2},\\
    Q_{ws}^{rad} &= \sigma A_{ws}\epsilon_w\epsilon_s\Omega(T^4_w-T^4_s),\quad 
    \Omega = \frac{L_c}{2(\pi-\psi)r_c}.
\end{align}
\end{subequations}
$\alpha_g$ is the absorptivity of the gasses.
The thermal radiation in the axial direction is assumed to be negligible.

\paragraph{Emissivity and absorptivity}
The emissivity of the solid bulk, refractory, and wall of the kiln is reported by \cite{Hanein2017} as $\epsilon_s=0.9$, $\epsilon_r=0.85$, and  $\epsilon_w=0.8$.
The emissivity of the refractories and walls of each module are assumed to be identical to the kiln emissivities.  
The emissivity of the solid-gas mixture is inspired by \cite{ALBERTI2018274}:
\begin{align}
    \epsilon_m = \epsilon_s + \epsilon_g - \Delta\epsilon^s_g, \quad \Delta\epsilon^s_g = \epsilon_s\epsilon_g,
\end{align}
where $\Delta\epsilon^s_g$ is the overlap emissivity. The particle emissivity is assumed identical to the solid bulk emissivity of the kiln, $\epsilon_p=\epsilon_s$.

The emissivity of gas, $\epsilon_g$, can be computed according to the WSGG model of 4 grey gases \citep{Johanson2011}:
\begin{subequations}
\begin{align}
    \epsilon_g &= \sum^4_{j=0}a_j(1-e^{-k_jS_mP(x_{H2O}+x_{CO2})}),\\
    a_0 &= 1-\sum^4_{j=1}a_j,\quad a_j=\sum^3_{i=1}c_{j,i}(\frac{T}{T_{ref}})^{i-1},\\
    k_j &= K_{1,j}+K_{2,j}\frac{x_{H2O}}{x_{CO2}}, \quad k_0=0, \quad x_i = \frac{n_i}{n_g},\\
    c_{j,i} &= C_{1,j,i}+C_{2,j,i}\frac{x_{H2O}}{x_{CO2}}+C_{3,j,i}(\frac{x_{H2O}}{x_{CO2}})^2.
\end{align}
\end{subequations}
 This assumes only \ce{H2O} and \ce{CO2} affect the radiation, with other gasses being transparent. $T_{ref}$ is 1200 K. The $K$ and $C$ coefficients are given by the look-up table in \cite{Johanson2011}. $S_m$ is the average total path length.
The absorptivity of the gas can be defined as a function of the emissivity \citep{Perry}: 
\begin{align}
    \alpha_g = \epsilon_g(T_s)P_mS_m\sqrt{\frac{T_s}{T_g}}.
\end{align}
$P_m$ is the partial pressure of \ce{CO2} and \ce{H2O}. 
\cite{Gorog1981} reports the path length in the kiln to be $S_m = 0.95(2r_c-h)$.
For the cooler, $S_m$ is taken as the distance between the centroid of the phases, $h/2$. 
The path length of the calciner and risers is $S_m = \sqrt{A_{cr}/\pi}$.

\subsection{Transport phenomena}
The transport between segments is described by fluxes. The molar fluxes, $N$, are given by
\begin{align}
    N = N_{a} + N_{d}, \quad N_a = v C, \quad N_d = -D \nabla C,
\end{align}
where $N_a$ is the advective term and $N_d$ is the diffusion term given by Fick's law. The heat fluxes, $N_U$, are given by
\begin{align}
    N_U = \tilde{H} + \tilde{Q}, \quad \tilde{H} = H(T,P,N), \quad \tilde{Q} = -k\nabla T,
\end{align}
where $\tilde{H}$ is the enthalpy flux and $\tilde{Q}$ is the thermal conductivity given by Fourier's law.
The diffusion and conductivity in each module are only included in transport between segments internally in each module. Thus for a module with $n_i$ segments in the $i$th dimension, $i \in \{x,y,z\}$, the molar flux for the $j$th  segment is given by
\begin{align}
    N_{i,j+\frac{1}{2}} = \begin{cases}
        N_{in}, &j=0\\
        v_{i,j+\frac{1}{2}} C_{i,j+\frac{1}{2}} - D_{i,j+\frac{1}{2}}\partial_i C_{j+\frac{1}{2}}, & 1<j<n_i\\
        v_{i,n_i+\frac{1}{2}} C_{i,n_i+\frac{1}{2}}, &j= n_i\\
    \end{cases}.
\end{align}
$N_{in}$ is the influx and $N_{i,n_i+\frac{1}{2}}$ is the outflux. The partial concentration $\partial_iC_j$ is approximated as $\partial_iC_{j+\frac{1}{2}} = ( C_{j+1}-C_j )/\Delta i$.
In the advection term, the concentrations, $C_{i,j+\frac{1}{2}}$, are defined by the velocity as
\begin{align}
    C_{i,j+\frac{1}{2}} = \begin{cases}
        C_{i,j} &  v_{i,j+\frac{1}{2}} \geq 0\\
        C_{i,j+1}& v_{i,j+\frac{1}{2}} <0
    \end{cases}.
\end{align}
Assuming no back-flow between modules, the boundary concentration at the end, $C_{n_i+\frac{1}{2}}$, is given as
\begin{align}
    C_{i,n_i+\frac{1}{2}} = \begin{cases}
        C_{i,n_i}, & v_{i,n_i+\frac{1}{2}} \geq 0\\
        0 & v_{i,n_i+\frac{1}{2}} <0
    \end{cases}.
\end{align}
For the kiln gas phase, the orientation is flipped, thus the influx, $N_{in}$ is at $j=n_i$ instead, and the exit is at $j=1$, with boundary concentration as
\begin{align}
    C_{i,\frac{1}{2}} = \begin{cases}
        0, & v_{i,\frac{1}{2}} \geq 0\\
        C_{i,1}, & v_{i,\frac{1}{2}} <0
    \end{cases}.
\end{align}

\subsubsection{Velocity}
In the kiln, the solid bulk moves down the kiln in a cascading motion due to the kiln rotation, $\omega$.
According to \cite{Saeman1951}, the average velocity of the bulk is
\begin{subequations}
\begin{align}
    v_{s}(z) &= \omega \frac{\psi+\phi(z) \text{cos}(\xi)}{\text{sin}(\xi)}\frac{\pi L_c(z) }{\text{asin}(\frac{L_c(z)}{2r_c})},\\
     \xi &= a_\omega\omega + b_\omega.
\end{align}
\end{subequations}
$\xi$ is the repose angle, with coefficients determined experimentally \citep{Yang2003,Yamane1998}.

The horizontal velocity of the solid bulk in the cooler is given by the grate belt carrying the clinkers.
We assume no vertical movement of the bulk, i.e., that no dust is carried off by the airflow. The velocities are thus,
\begin{align}
    v_{s}(z) = [v_{s,y},v_{s,z}]^T = [0,v_{grate}(z)]^T.
\end{align}
In the calciner and risers, the solid particles are assumed to move with the velocity of the gas.
 
As the kiln gas velocity is below 0.2 Mach speed \citep{MortenPHD}, then according to \cite{Darcy-Howel}, the Darcy-Weisbach equation for incompressible gas is a valid description using a turbulent flow Darcy friction factor $f_D$,
\begin{subequations}
\begin{align} 
    v_{z,g}^2 = 2\frac{|\Delta P|}{\Delta z}\frac{D_H}{f_D\rho_g},\ f_D &= \frac{0.316f_{0}}{\sqrt[4]{Re}} = \frac{0.316f_{0}\sqrt[4]{\mu_g}}{\sqrt[4]{\rho_g |v_{z,g}|D_H}}, \\ D_H &= \frac{4V_g}{A_{gw}+A_{gs}}.
\end{align}
\end{subequations}
$D_H$ is the hydraulic diameter for a Non-uniform nor circular channel \cite{HESSELGREAVES20171}. $f_0$ is a correction factor we introduce for empirical tuning. The gas velocity can then be formulated as
\begin{align} 
    v_{z,g} = \Bigg(\frac{2}{0.316f_0}\sqrt[4]{\frac{D_H^{5}}{\mu_g\rho_g^3}}\frac{|\Delta P|}{\Delta z}\Bigg)^{\frac{4}{7}}\text{sgn}\Big(-\frac{\Delta P}{\Delta z}\Big)\label{eq:vel}.
\end{align}
$\text{sgn}\Big(-\frac{\Delta P}{\Delta z}\Big)$ represents the velocity sign convention, where $v_{z,g}<0$ for the gas flowing up the kiln.

For the calciner, risers, and cooler the gas velocity is assumed to be described by the same Darcy-Weisbach equation. For the calciner and risers, the velocity, $v_{y,m}$, is formulated for the $y$-direction using \eqref{eq:vel}  with the mixture viscosity, $\mu_m$, mixture density, $\bar\rho_m$, and hydraulic diameter $D_H =  \frac{4V_\Delta}{A_{c}}$.

The gas velocity in the cooler is both horizontal and vertical,
\begin{align}
    v_{g} &= [v_{y,g},v_{z,g}]^T.
\end{align}
The hydraulic diameters are given as
\begin{subequations}
\begin{align}
   D_{H,y} &= \frac{4V_g}{2 A_{yz}+2 A_{wy}+A_{cl}},\ A_{wy}= w\Delta y,\\
   D_{H,z} &= \frac{4V_g}{2 A_{yz}+2 A_{wz}+A_{cl}}, \  A_{wz}= w\Delta z,\\
   A_{yz} &= \Delta z\Delta y, \  A_{cl} = \frac{V_s}{\frac{\pi}{6} D_p^3} (\pi D_p^2) = \frac{6V_s}{D_p}.
\end{align}
\end{subequations}
$A_{cl}$ is the total surface area of the clinkers.
\subsubsection{Cyclone flux and efficiency}
The flux of the cyclone model is described as
\begin{subequations}
\begin{align}
    N_{s,in,i} &= (1-\eta_{sal})v_{in}C_{s,in,i}, \quad &N_{g,in,i} &= v_{in}C_{g,in,i},\\
    N_{s,x,i} &= v_{s,x}C_{s,i}, \quad &N_{g,x,i} &= v_{g,x}C_{g,i},\\
    N_{s,sep,i} &= v_{s,sep}C_{s,i}, \quad &N_{s,sal,i} &= \eta_{sal}v_{in}C_{s,in,i}.
\end{align}
\end{subequations}
$N_{s,in}$ and $N_{g,in}$ are the fluxes interring the chamber of the cyclones. $N_{s, sal}$ is the flux of solids separated by saltation, i.e., particles separated on entry by flying into the wall. $\eta_{sal}$ is the saltation separation efficiency.

The outlet gas velocity, $v_{g,x}$, of the exit gas flux, $N_{g,x}$, is described by the turbulent Darcy-Weisbach equation in \eqref{eq:vel} over the distance $\Delta y = h_x$ with the mixture viscosity, $\mu_m$, mixture density, $\bar\rho_m$, and hydraulic diameter $D_H = d_x$. Thus the pressure located below the outlet pipe is assumed representative of the pressure in the entire cyclone.

The outlet solid velocity, $v_{s,x}$, of the exit solid flux, $N_{g,x}$, is assumed to correlate to the outlet gas velocity, $v_{g,x}$:
\begin{align}\label{eq:vsout}
    v_{s,x} = f_N v_{g,x}.
\end{align}
$f_N$ is a correction factor for tuning efficiency.

The separation flux, $N_{s, sep}$, is the solid particles separated by hitting the cyclone wall by the swirling motion of the solid-gas mixture. The separation velocity, $v_{sep}$, is the radial particle velocity at the wall:
\begin{align}
    v_{sep} = \frac{d_m^2\Delta\rho}{18\mu_m}\frac{v_{\theta,r_{eq}}^2}{r_{eq}}, \quad \Delta\rho = \rho_{s} -\rho_g, \quad r_{eq} = \sqrt{\frac{V_{\Delta}}{\pi h_t}}. 
\end{align}
$d_m$ is the median particle diameter. $\Delta\rho$ is the difference between the solid particle density, $\rho_{s}$, and the gas density, $\rho_g$. $v_{\theta,r_{eq}}$ is the tangential velocity at radius $r_{eq}$.
The formulation is based on the approach of Mothes and Löffler \citep{Hoffmann07}, approximating the cyclone by a cylinder of equivalent volume with radius $r_{eq}$.

The formulation for the tangential velocity of a given radius $r$ is given by Muschelknautz \citep{Hoffmann07}:
\begin{align}
    v_{\theta,r} &= \frac{ \frac{r_{c}}{r}v_{\theta,w}}{(1+\frac{f_SA_{sep}v_{\theta,w}}{2A_{in}v_{in}} \sqrt{\frac{r_c}{r}}) }, \quad f_S = 0.005 (1 + 3\sqrt{c_0}).
\end{align}
$f_S$ is the drag friction factor. $v_{\theta,w}$ is the inlet tangential velocity at the wall, formulated as
\begin{subequations}
\begin{alignat}{2}
 v_{\theta,w} & = \frac{r_{in}}{r_{c}\alpha}v_{in},\quad\beta = \frac{w_{in}}{r_c}, \quad c_0 = \frac{\sum_iM_iC_{s,in,i}}{\sum_jM_jC_{g,in,j}},\\
    \alpha & = \frac{1 - \sqrt{1 - \beta(2 - \beta)\sqrt{1 - \beta(2-\beta)\frac{1-\beta^2}{1+c_0} } }}{\beta}.
\end{alignat}
\end{subequations} 
$\alpha$ is the inlet constriction coefficient. $c_0$ is the inlet load ratio.

The total efficiency of the separation, $\eta$, consists of the effect of saltation and internal vortex separation \citep{Hoffmann07}:
\begin{equation}
    \eta = \eta_{sal} + \eta_{sep} = \frac{\dot m_{s,sal}+\dot m_{s,sep}}{\dot m_{s,in}}.
\end{equation}
The separation efficiency, $\eta_{sep}$, depends on the relative separation and outlet velocities.
The saltation efficiency, $\eta_{sal}$, is formulated as
\begin{equation}
    \eta_{sal} = 1 - \min\bigg(1,\frac{c_{0L}}{c_0}\bigg).
\end{equation}
$c_{0L}$ is the cyclone loading limit,
\begin{subequations}
\begin{align}\label{eq:col}
    c_{0L} &= f_c\cdot 0.025\bigg(\frac{d^*}{d_{m}}\bigg)(10c_0)^k, \\
    k &= 0.15\delta + (-0.11-0.10\text{ln}(c_0))(1-\delta),\\
    \delta &= 1 \Leftrightarrow c_0 \geq 0.1.
\end{align}
\end{subequations}
$f_c$ is a correction factor for tuning efficiency and density. $d^*$ is the particle cut-size,
\begin{align}    
    d^* &= \sqrt{\frac{18\mu_m0.9A_{in}v_{in}}{\Delta\rho 2\pi h_i v_{\theta,r_x}^2}}.
\end{align}
$v_{\theta,r_x}$ is the tangential velocity at the outlet radius $r_x$. $h_i$ is the height of the cyclone below the outlet pipe, $ h_i  = h_t - h_{x}$.
The $0.9$ factor corresponds to an assumed $10\%$ gas flow from the inlet directly to the outlet area.

\subsubsection{Diffusion}
The diffusion term is assumed zero in all modules except the kiln. The diffusion across segments within the solid phase of the kiln is assumed zero, as it is negligible according to \cite{Mujumdar2006} since an industrial kiln has a Peclet number greater than $10^4$.
For the diffusion within the kiln gas, the coefficient, $D_{g, i}$, of compound $i$ is given by the correlation of Fuller's model \citep{Poling2001Book}:
\begin{subequations}
\begin{align}    
     D_{g,i} & = \bigg(\sum_{\substack{j=1\\j\neq i}}\frac{x_j}{c_gD_{ij}}\bigg)^{-1},\quad x_j = \frac{C_{g,j}}{c_g}\label{eq:gdiff}, \quad c_g= \sum_jC_{g,j},\\
    D_{i,j} &= \frac{0.00143T^{1.75}}{P M_{ij}^{\frac{1}{2}}[(V_{\sum})^{\frac{1}{3}}_i+(V_{\sum})^{\frac{1}{3}}_j]^2}\label{eq:gdiff2}, \quad M_{ij} = \frac{2}{\frac{1}{M_i}+\frac{1}{M_j}}. 
\end{align}
\end{subequations}
$x_j$ is the mole fraction. $V_{\sum}$ is the diffusion volume. $M$ is the molar mass.

\subsubsection{Enthalpy flux and conductivity}
The formulation of the heat transport depends on the phases of the module.
For the kiln and cooler modules, the enthalpy flux 
 and thermal conduction is given for the solid, gas, refractory, and wall phases:
 \begin{subequations}
\begin{align}
    \Tilde{H}_i &= H(T_i,P,N_i), \quad \text{for }i=\{s,g\},\\
    \Tilde{Q}_i &= -k_i\nabla{T_i}, \quad \text{for }i=\{s,g,r,w\}.
\end{align}
\end{subequations}
For the cyclones, calciner, and risers, the formulation is for the mixture, refractory, wall phases:
\begin{subequations}
\begin{align}
    \Tilde{H}_m &= H(T_m,P,N_g) + H(T_m,P,N_s),\\
    \Tilde{Q}_i &= -k_i\nabla{T}_i, \quad \text{for }i=\{m,r,w\}.
\end{align}
\end{subequations}

\subsection{Fallthrough flow}
Fall-through flows are flows in the risers that exit down through the eye of the vortex of the cyclone below. 
Assuming a given riser has a fixed loss to fall-through, a percentage $p$, the fall-through flux, $N_{s, fall}$, is proportional to the influxes from the cyclone below and above. The corresponding heat flow depends on the riser conditions and the fall-through flux,
\begin{subequations}
\begin{align}
    N_{s,fall} &= p N_{s,in} = p ( N_{s,up} + N_{s,d}),\\
    \tilde H_{s,fall} &= H(T_m,P,N_{s,fall}).
\end{align}
\end{subequations}

In the cyclone, the fall-trough flow is assumed to go straight through the cyclone. Thus, the entire fall-through flow exits the cyclone at the bottom together with the material saltation fluxes, $N_{s, sal}$, and separation fluxes, $N_{s, sep}$:
\begin{subequations}
\begin{align}
    N_d &= \frac{A_{in}}{A_d}N_{s,sal} + \frac{A_{sep}}{A_d}N_{s,sep} + \frac{A_x}{A_d}N_{s,fall},\\
    \tilde H_d &= \frac{A_{in}}{A_d}\tilde H_{s,sal} + \frac{A_{sep}}{A_d}\tilde H_{s,sep} + \frac{A_x}{A_d}\tilde H_{s,fall}.
\end{align}
\end{subequations}

\begin{table*}[ht]
    \centering
    \caption{Each module connects to specific points in the other modules.}
    \begin{footnotesize}  
    \begin{tabular}{c|c|c|c|c|c|c|c|c|c|c|c|c|c}\hline
     To$\backslash$From     & $Cy1$ & $R1$ &$Cy2$ &$R2$ &$Cy3$ &$R3$ &$Cy4$ &$R4$ &$Cy5$& $Ca$ &$K$ &$R5$ &$Co$ \\ \hline
    $Cy1$ & - & $\frac{A_{out,r1}}{A_{in,cy1}}$ & - & - & - & - & - & - & - & - & - & - & -  \\
    $R1$  & - & - & $\frac{A_{x,cy2}}{A_{in,r1}}$ & - & - & - & - & - & - & - & - & - & -  \\
    $Cy2$ & - & $\frac{A_{in,r1}}{A_{x,cy2}}$ & - & $\frac{A_{out,r2}}{A_{in,cy2}}$ & - & - & - & - & - & - & - & - & - \\
    $R2$  & $\frac{A_{d,cy1}}{A_{in,r2}}$ & - & - & - &$\frac{A_{x,cy3}}{A_{in,r2}}$ & - & - & - & - & - & - & - & - \\
    $Cy3$ & - & - & - &$\frac{A_{in,r2}}{A_{x,cy3}}$ & - &$\frac{A_{out,r3}}{A_{in,cy3}}$ & - & - & - & - & - & - & - \\
    $R3$  & - & - & $\frac{A_{d,cy2}}{A_{in,r3}}$ & - & - & - & $\frac{A_{x,cy4}}{A_{in,r3}}$ & - & - & - & - & - & -  \\
    $Cy4$ & - & - & - & - & - & $\frac{A_{in,r3}}{A_{x,cy4}}$ & - & $\frac{A_{out,r4}}{A_{in,cy4}}$ & - & - & - & - & -   \\
    $R4$  & - & - & - & - & $\frac{A_{d,cy3}}{A_{in,r4}}$ & - & - & - & $\frac{A_{x,cy5}}{A_{in,r4}}$ & - & - & - & -   \\
    $Cy5$ & - & - & - & - & - & - & - & $\frac{A_{in,r4}}{A_{x,cy5}}$ & - & $\frac{A_{out,ca}}{A_{in,cy5}}$ & - & - & -  \\
    $Ca$  & - & - & - & - & - & - & $\frac{A_{d,cy4}}{A_{in,ca}}$ & - & - & - & $\frac{A_{in,K}}{A_{in,ca}}$ & $\frac{A_{out,r5}}{A_{in,ca}}$ & -  \\
    $K$   & - & - & - & - & - & - & - & - & $\frac{A_{d,cy5}}{A_{in,K}}$ & - & - & - & $\frac{A_{y1,co}}{A_{out,K}}$  \\
    $R5$  & - & - & - & - & - & - & - & - & - & - & - & - & $\frac{A_{y2,co}}{A_{in,r5}}$  \\
    $Co$  & - & - & - & - & - & - & - & - & - & - & $\frac{A_{out,K}}{A_{z,in,co}}$ & - & -  \\ \hline
    \end{tabular}   
    \end{footnotesize}
    \label{tab:connection}
\end{table*}
\subsection{Module connections}
Each module is connected by its boundary fluxes, such that the outflux of one module is the influx of another.
The general idea is expressed by
\begin{align}
    N_{in,j} = \frac{A_{out,i}}{A_{in,j}}N_{out,i}.
\end{align}
The connection enthalpy flux is likewise given by
\begin{align}
    \Tilde{H}_{in,j} = \frac{A_{out,i}}{A_{in,j}}\Tilde{H}_{out,i}.
\end{align}
Table \ref{tab:connection} shows the connecting area ratios of each module connection.
The roof across the cooler is split into four parts (sequential $A_{y1, co}$, $A_{y2, co}$, $A_{y3, co}$, and $A_{y4, co}$) of varying sizes. $A_{y1,co}$ connects the gas outflow to the kiln. $A_{y2,co}$ connects to the tertiary pipe. $A_{y3, co}$ denotes the part with no exit for the gas. $A_{y4,co}$ connects to the environment. The pressure above the cooler, $P_{y, co}$, is then the sum of neighboring pressures scaled with their connection area:
\begin{subequations}
\begin{align}
    P_{y,co} &= \sum P_{i,co},\ P_{i,co} = \frac{A_{i,co}}{A_{zw}}P_{i},  \quad i\in \{y1,y2,y3,y4\},\\
    P_{y1} &= P_{out,K}, \ P_{y2} = P_{in,r5}, \ P_{y3} = P_{co}, \ P_{y4} = P_{env}.
\end{align}
\end{subequations}
$A_{zw}$ is the roof cross-area of a cell, while $A_{i,co}$ is that cell's area connecting to each module.

\subsection{Boundary conditions}
The full model has the following boundary conditions:
\begin{itemize}
    \item The fan-induced pressure, $P_{out}$, above cyclone 1.
    \item The molar influx, $N_{s, in}$, and temperature, $T_{s, in}$, of the material feed interring at riser 1.
    \item The fuel influxes, $N_{fuel, i}$, and temperatures, $T_{fuel, i}$ of the calciner and kiln, $i\in\{Ca, K\}$.
    \item The primary air influx, $N_{pair}$, and temperature, $T_{pair}$.
    \item The kiln rotation velocity, $\omega$.
    \item The cooler grate velocities, $v_{grate}$.
    \item The cooler air influxes, $N_{air, co}$, and temperatures, $T_{air, co}$.
    \item  The temperature, pressure, and composition of the ambient air in the environment.
\end{itemize}

Further, the model has a potential boundary condition in the form of false air inflows.
False air is air creeping in due to the pressure difference between the environment and the internal part of the process. False air typically occurs at connections between structures; in the model, this corresponds to where the modules connect. At each module, the false air can be included as
\begin{subequations}
\begin{align}
    N_{in} &= N_{in} + N_{false}, \\
    H_{in} &= H_{in} + H(T_e,P_e,N_{false}).
\end{align}
\end{subequations}

%% file: Simulations.tex
\section{Simulation}
\label{sec:SimulationResults}
In this section, we present the results of a simulation of the model.
We consider the system from Fig. \ref{fig:process} with the dimensions given in Table \ref{tab:dimendion}.
Using the finite-volume approach, the calciner and each riser module are formulated using 3 volumes, the kiln by 10 volumes, and the cooler by 5 volumes.
We tune the model to match a scenario from FLSmidth Cement's database.
\begin{table*}
    \centering
    \caption{Module dimensions and false air.}
    \begin{footnotesize}
        \begin{tabular}{cc|cccccccccccc|c}\hline
          & Dimensions&  $h_t$ &  $h_c$ &  $h_x$ &  $A_{in}$ &  $r_c$ &$r_r$ &$r_w$ &$r_d$ &  $r_x$ &  $r_{in}$ &  $w_{in}$&   $l_{in}$ & False air\\ \hline
         Module &Units & m & m & m & m$^2$ & m & m & m & m & m & m &  m& m & m$^3$/s\\ \hline 
         cyclone & $Cy1$& 18.28  &  7.43  &  3.45  &  10.95 &  3.45  &3.59 &3.60  &0.32  &  1.92  &  2.76  &  1.38 & 0.1 & 0.92\\
         cyclone &$Cy2$& 11.38  &  7.28  &  3.41  &  13.34 &  3.41  &3.59  &3.60  &0.45  &  2.43  &  2.73  &  1.37 &  0.1 & 0.90\\
         cyclone &$Cy3$& 11.24  &  7.79  &  3.38 &  13.70 &  3.38  &3.59  &3.60  &0.45 &  2.48  &  2.70  &  1.35 &  0.1 & 0.44\\
         cyclone &$Cy4$& 11.98  &  8.13  &  3.53  &  14.83 & 3.51  &3.74  &3.75  &0.45 &  2.58  &  2.82  &  1.41& 0.1 & 0.44\\
         cyclone &$Cy5$& 11.94  &  8.09  &  3.54  &  14.83 & 3.54&3.74 &3.75 &0.45 &  2.58 &  2.83  &  1.42 & 0.1 & 0.44\\ \hline \hline
         & Dimensions&  $L$ &  $r_c$ &$r_r$ & $\psi$& &&&&&&&&False air\\ \hline
        Module & Units & m & m & m & $^\circ$& &&&&&&&& m$^3$/s\\ \hline 
         kiln & $K$ & 51 &  1.98 &2.18 & 2& &&&&&&&&0.83\\  \hline  \hline
          &  Dimensions&  $h_{tot}$&  $h_{cl}$&  $h_{cu}$ &  $r_c$ &$r_r$&$r_w$&$r_l$&$r_u$ & $p_{fall}$& &&&False air\\ \hline
          Module&Units & m & m & m & m & m & m & m &m &\%& &&&m$^3$/s\\ \hline 
         calciner&$Ca$ & 33.00 &  33.00 &  0  &  3.08  &3.29  &3.30 &3.08 &3.08  & 0.0& &&&3.22\\
         riser &$R1$&  21.38 & 4.83&  0  &  2.45  &2.47 &2.48 &2.45 &1.72 & 0.1& &&&0.45\\
         riser &$R2$&  19.38 &4.18 &  0  &  2.45  &2.47 &2.48 &2.45 &1.90 & 0.1& &&&0.45\\
         riser &$R3$&  20.37 &4.18 &  0  &  2.55  &2.57 &2.58 &2.45 &1.90 & 0.1& &&&0.44\\
         riser &$R4$&  19.87 &4.37 &  0  &  2.55  &2.57 &2.58 &2.45 &1.99 & 0.0& &&&0.44\\
         riser &$R5$&  61.00 &61.00&  0  &  1.13  &1.33 &1.34 &1.13 &1.13 & 0.0& &&&0.0\\  \hline  \hline
         & Dimensions&  $L$ &  $h$ &$w$ & $A_{y1,co}$ & $A_{y2,co}$ & $A_{y3,co}$& $A_{y4,co}$& &&&&&False air\\  \hline
         Module &  Units & m & m & m & m$^2$& m$^2$& m$^2$ & m$^2$& &&&&&m$^3$/s\\ \hline          
         cooler&$Co$    & 36& 3 &4  & 12.1 & 22.5 & 49.0  & 60.5m& &&&&&0.0\\  \hline
    \end{tabular}
    \end{footnotesize}  
    \label{tab:dimendion}
\end{table*}
\begin{table*}[]
    \centering
    \caption{Composition of scenario flows: fuels, primary air, and meal flows.}
    \begin{footnotesize}  
    \begin{tabular}{c|ccccc|cccc} \hline
         Compound &  \ce{C} & \ce{CaO} & \ce{SiO2} & \ce{Al_2O_3} & \ce{Fe_2O_3} & \ce{N2} & \ce{O_2}&\ce{Ar} & \ce{H2O}\\ \hline
        Units & kg/s & kg/s&kg/s&kg/s&kg/s&kg/s&kg/s& kg/s& kg/s \\ \hline
        Calciner fuel & 4.46 & 0.01&0.37&0.23&0.025&1.50&2.16&0.03& 0.02\\
        Kiln fuel & 1.98 & 0.00&0.05&0.03&0.00&1.10&0.34&0.02& 0.00\\ 
        primary air &0.00 &0.00&0.00&0.00&0.00&1.51&0.46&0.026&0.01\\ \hline \hline
        Compound &  \ce{CaCO3} & \ce{CaO} & \ce{SiO2} & \ce{Al_2O_3} & \ce{Fe_2O_3}& Metakaolin &\ce{SiO2}& \ce{Al_2O_3}\\ \hline
        Units & kg/s& kg/s& kg/s & kg/s&kg/s&kg/s &kg/s&kg/s\\ \hline
        meal flow & 69.5 &  0.0&7.5&0.1&1.7& 6.9 & 3.7 & 3.2\\ \hline
    \end{tabular}    
    \end{footnotesize}
    \label{tab:fuel}
\end{table*}

\subsection{Scenario}
In the scenario, the material meal feed flow is 85.7 kg/s at 60$^\circ$C. The calciner fuel is injected at 8.94 kg/s and 63.9$^\circ$C. 
The kiln fuel and primary air are injected at 3.57 kg/s and 2.01 kg/s at 43.8$^\circ$C.
Table \ref{tab:fuel} shows the compositions of the meal, fuels, and primary air. The metakaolin (\ce{Al_2O_3*2SiO2}) is treated as its constituency, \ce{SiO_2} and \ce{Al_2O_3}.

For the scenario, the kiln rotates with 4 rotations per minute, $\omega = 4$ rpm. The grate belts in the cooler have a 36-minute retention time, $v_{grate}=0.017$ m/s.
The total cooler air flow is 157.9 m$^3$/s allocated on each segment by 28.0\% on the first and 27.0\% on the rest. 
The induced outlet pressure, $P_{out}$, is 0.9452 bar above the preheating tower.

The environment has the ambient temperature, $T_e$, is 25$^\circ$C and the ambient pressure, $P_e$, is 1 bar.
The false air flows received by each module in the scenario are provided in the last column of table \ref{tab:dimendion}.
The cyclones 1 to 5 in the scenario have a steady-state efficiency of 94.9\%, 89\%, 87\%, 85\%, and 75\%, respectively.

\subsection{Tuning}
To fit the model to the scenario, the following variables are tuned.
The correction factors, $f_0$, are tuned for the pressure drop of each module as 575, 1280, 1380, 1430, and 350 for the cyclones, 78, 55, 65, 60, and 55 for the risers, 1350 for the calciner, 2 for the kiln, and 100 for the cooler.

Table \ref{tab:tuneCyclone} shows the tuned correction factors of the cyclones, reported in \cite{Svensen2024Cyclone} for the individual tuning of each cyclone.
  \begin{table}
    \centering
      \caption{Correction factors for cyclone tuning.}
      \begin{footnotesize}  
      \begin{tabular}{c|ccccc}\hline
                & $Cy_1$ & $Cy_2$ & $Cy_3$ & $Cy_4$ & $Cy_5$\\ \hline
          $f_c$ & 6.5&4.2&4.85&5.2&6.72\\
          $f_N$ & 1/22&1/10.1&1/8.5&1/7.3&1/4.2\\ \hline
      \end{tabular}
      \end{footnotesize}
      \label{tab:tuneCyclone}
  \end{table}
In the kiln, the porosity is tuned to fit the nominal fill ratio and nominal throughput for the dimensions, $\phi_p=8/9$.
To correct the heat transfer between phases in the kiln to the scenario, a correction factor is introduced in \eqref{eq:betagsKiln}, $\beta_{gs} := 3\beta_{gs}$.

The reaction rate coefficients, $k_0$, of reaction $r_1$ to $r_{12}$ in section \ref{secsub:kinetic} are tuned using Table \ref{tab:allreactions} as base values. $r_1$ is scaled by $170$. $r_2$ is scaled by $200\cdot 10^3$. $r_3$ is scaled by $60\cdot 10^5$. $r_4$ is scaled by $5\cdot 10^6$. $r_5$ is scaled by $5\cdot 10^{10}$. $r_7$ is scaled by $5\cdot 10^5$. $r_{10}$ is scaled by $60$.

\subsection{Simulation}
A simulation of 35 hours of clinker production is used to demonstrate the model. The simulation is started in the steady state of the scenario. After 5 hours, the feed includes a step increase of 1 kg/s metakaolin, 1.17\% of total meal flow.
This particular step is chosen since the base scenario has a free lime content of 2.87\%, while a content around 1\% is typically the desired upper limit.

\subsubsection{Dynamic performance}
\begin{figure*}
    \begin{subfigure}[b]{\textwidth}
       \centering
        \includegraphics[width=\textwidth]{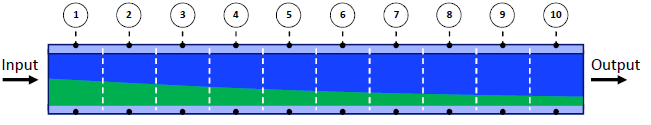}
         \caption{Kiln partitioning}
    \end{subfigure}
    \begin{subfigure}[b]{\textwidth}
       \centering
         \includegraphics[width=0.82\textwidth]{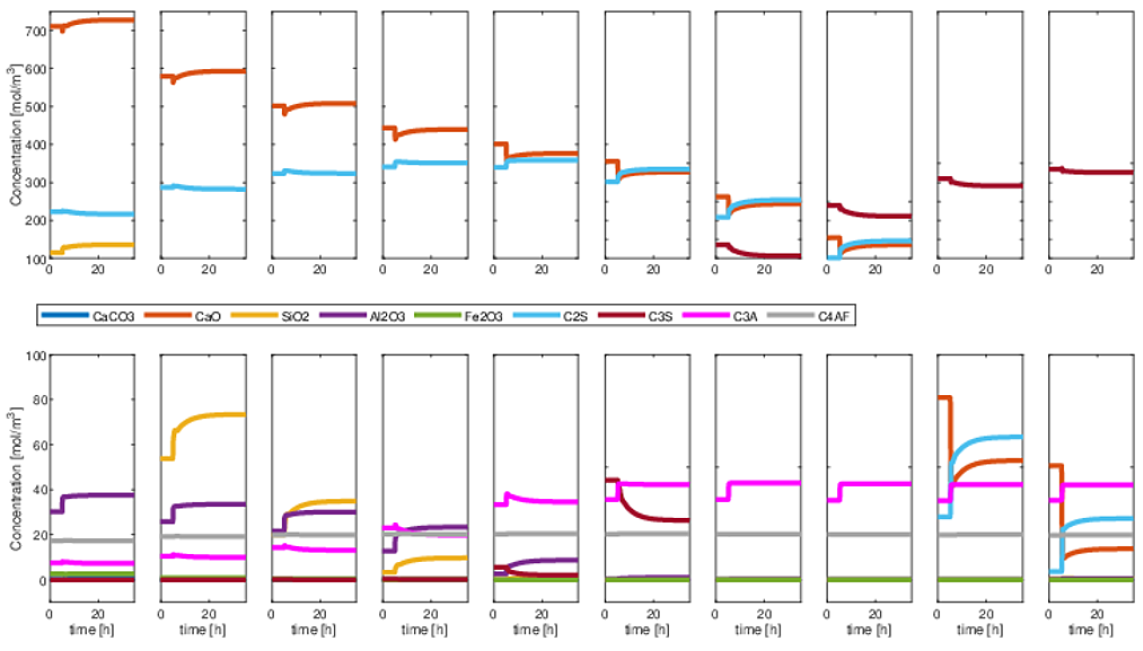}
    \caption{The evolution of solid material concentrations along the kiln. The kiln is split into ten segments. }
    \label{fig:solidCon}
    \end{subfigure}
    \begin{subfigure}[b]{\textwidth}
       \centering
         \includegraphics[width=0.82\textwidth]{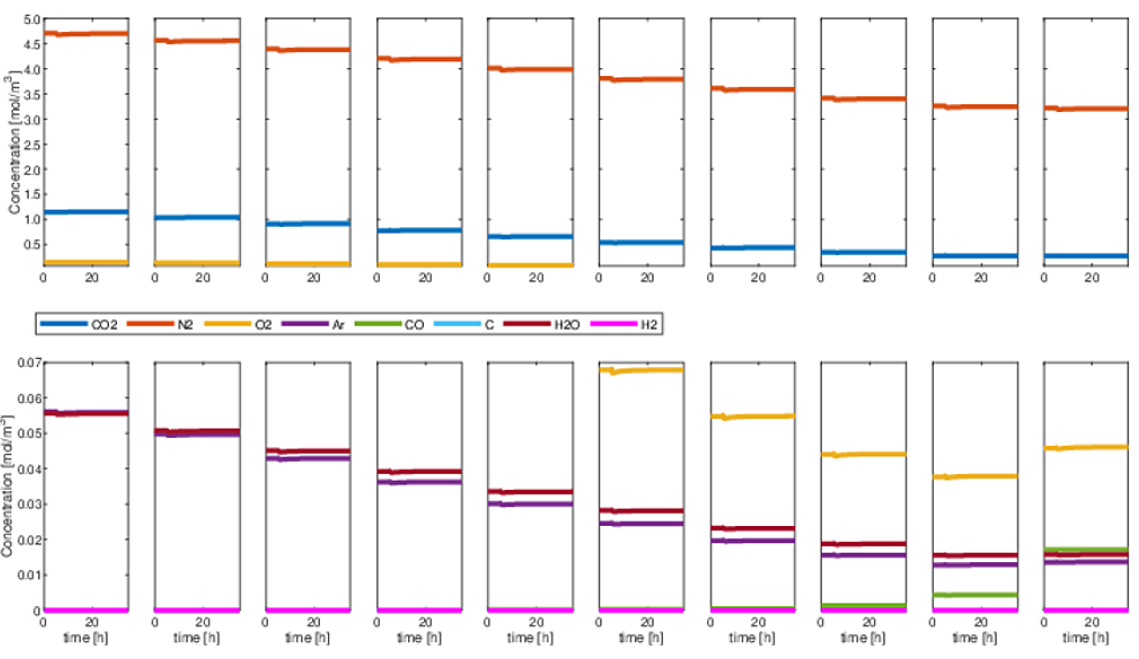}
    \caption{The evolution of gas concentrations along the kiln. The kiln is split into ten segments. }
    \label{fig:gasCon}
    \end{subfigure}
     \caption{Dynamic evolution of concentrations in the kiln module. For visibility, each segment and phase is shown in two concentration ranges.}
    \label{fig:kilnCon}
\end{figure*}

\begin{figure*}
       \centering
         \includegraphics[width=0.75\textwidth]{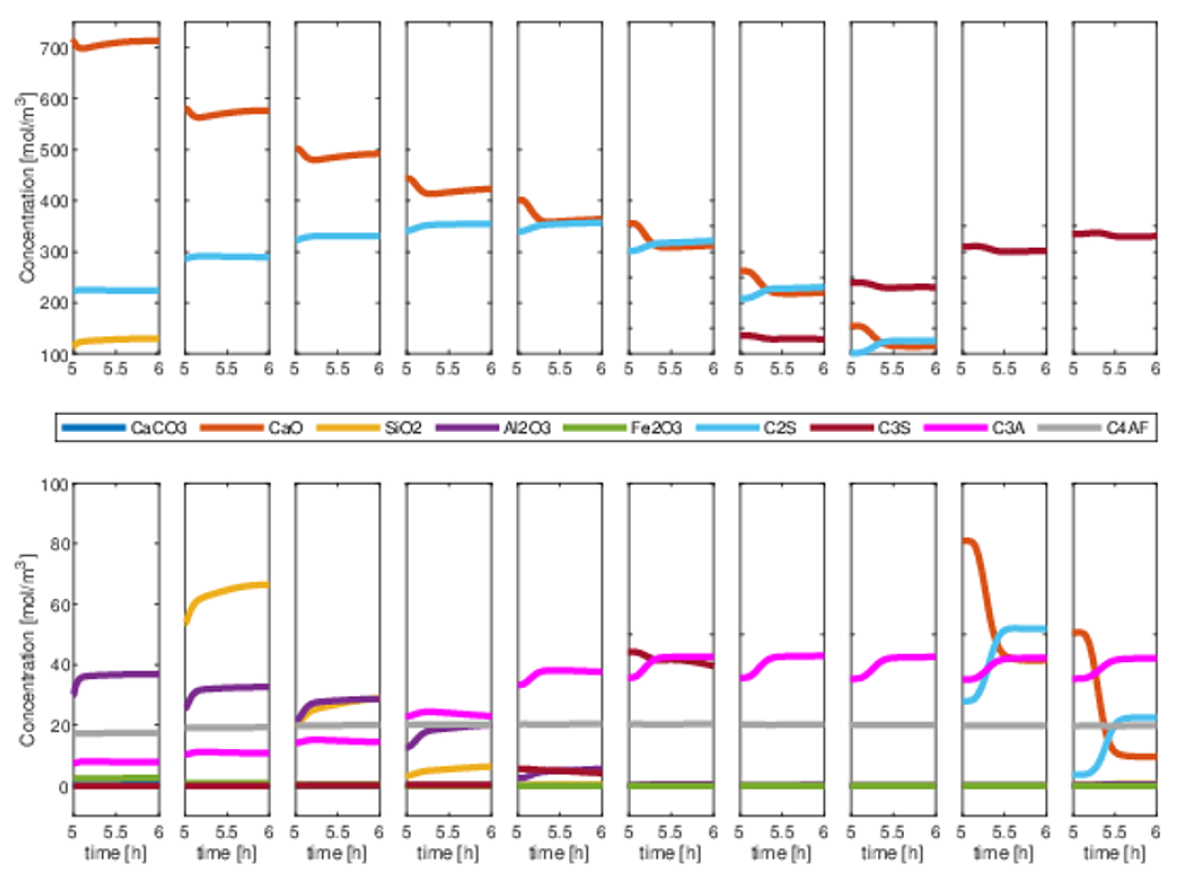}
     \caption{The changes in kiln solid concentrations during simulation hour 5-6 as a response to a step in meal intake at hour 5.}
    \label{fig:zoom}
\end{figure*}

Fig. \ref{fig:kilnCon} shows how the concentration of each compound changes over time in the kiln module. 
We can observe how fast the different compounds change across the kiln.
Fig. \ref{fig:zoom} shows the period immediately after the step, $t=5h$, with the response of each material starting in turn as the material moves down the kiln. We can see that after half an hour, the changes in the concentration shift from responding to the change in the feed flow directly to the change in feed composition and temperatures, driving the chemical reactions.
Similarly, Fig. \ref{fig:gasCon} shows the gas concentration with a fast response to the step and a following slow settling in response to volume changes.

\begin{figure*}[t]
    \centering
        \includegraphics[width=1.0\textwidth]{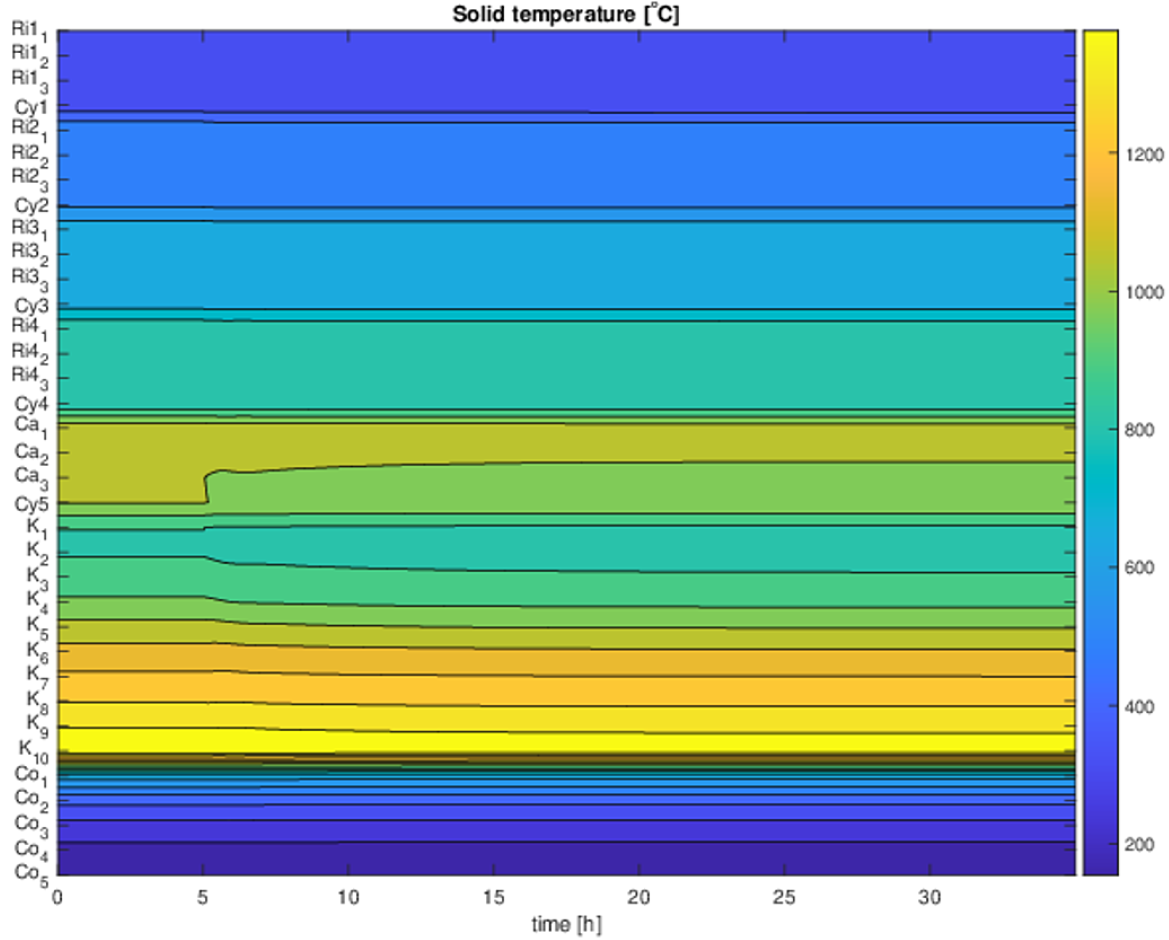}
         \caption{The temperature evolution of the solid materials across the pyro-process. Each module segment (y-axis) is ordered by the material path in the pyro-process, starting from the feed entry to the clinker exit.}
    \label{fig:TempDyn}
\end{figure*}

Fig. \ref{fig:TempDyn} shows the temperature of the solid materials across the pyro-process over time.
From the changes in the contour lines, we can observe that the overall system drops in temperature after the step, as expected with an increase in the feed-to-heat ratio.
This corresponds to the observed reactions in Fig. \ref{fig:solidCon}, where we see the \ce{CaO} concentration increase in the colder parts of the kiln, despite the feed step providing more \ce{SiO_2} and \ce{Al2O3} for reacting with the \ce{CaO}.
\begin{table*}[t]
    \centering
    \caption{Steady state composition of produced clinker (mass \%) and gas exhaust (vol. \%).}
    \begin{footnotesize}  
    \begin{tabular}{c|ccccccccc}\hline
       Clinker  & \ce{CaCO3}& \ce{CaO} & \ce{SiO2}& \ce{Al2O3}& \ce{Fe2O3}& C2S & C3S & C3A & C4AF\\ \hline
       Units & wt\%& wt\%& wt\%& wt\%& wt\%& wt\%& wt\%& wt\%& wt\%\\ \hline
      start   & 0.00 & 2.87 & 0.00 & 0.01 & 0.01 & 0.63 & 77.12 &  9.65 & 9.71\\
      end     & 0.00 & 0.76 & 0.03 & 0.05 & 0.01 & 4.64 & 73.67 & 11.28 & 9.56 \\ \hline \hline
      Gas Exhaust & \ce{CO2} & \ce{N2} & \ce{O2} & \ce{Ar} & \ce{CO} & \ce{C} & \ce{H2O} & \ce{H2}  \\ \hline
      Units & vol\%& vol\%& vol\%& vol\%& vol\%& vol\%& vol\%& vol\%\\ \hline
      start & 37.29 & 60.14 & 1.02 & 0.72 & 0.03 & 0.00 & 0.79 & 0.00\\
      end   & 37.27 & 60.15 & 1.04 & 0.72 & 0.03 & 0.00 & 0.79 & 0.00\\ \hline      
    \end{tabular}
    \end{footnotesize}
    \label{tab:steady}
\end{table*}
\subsubsection{Steady-state comparison}
Comparing the steady-state composition at the start and end of the simulation allows us to evaluate if the simulator performs as expected. 
Table \ref{tab:steady} shows the mass composition of the clinker and the volume composition of the exhaust gas at the start and end of the simulation. We observe that the free lime content is now 0.76\% instead of 2.87\% aligning with the intention for the feed step. Consequently, the C2S and C3A content has increased by 4\% and 1.6\%.
In addition, we can see that the content of C3S and C4AF has decreased by 3.5\% and 0.15\%, an indicator of the reduced temperature in the kiln. Minor changes are seen in the gas composition, e.g., \ce{CO2} has decreased by 0.02\%, indicating reduced combustion as \ce{CaCO3} is fully calcinated.
We observe that \ce{CO2} is 37\% of the exhaust volume, fitting the around 35\% observed in practice.

\begin{figure}[ht!]
    \centering
    \includegraphics[width=0.5\textwidth]{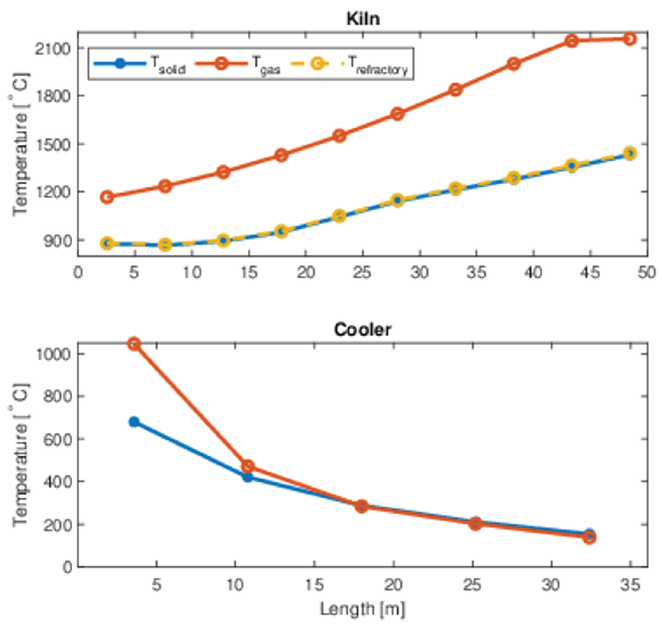}
    \caption{The steady-state temperature profiles of the kiln and the cooler.}
    \label{fig:tempKilnCooler}
\end{figure}
\begin{figure*}[ht!]
    \centering
    \includegraphics[width=0.9\textwidth,trim={2.0cm, 2.6cm, 3.4cm, 1.5cm},clip]{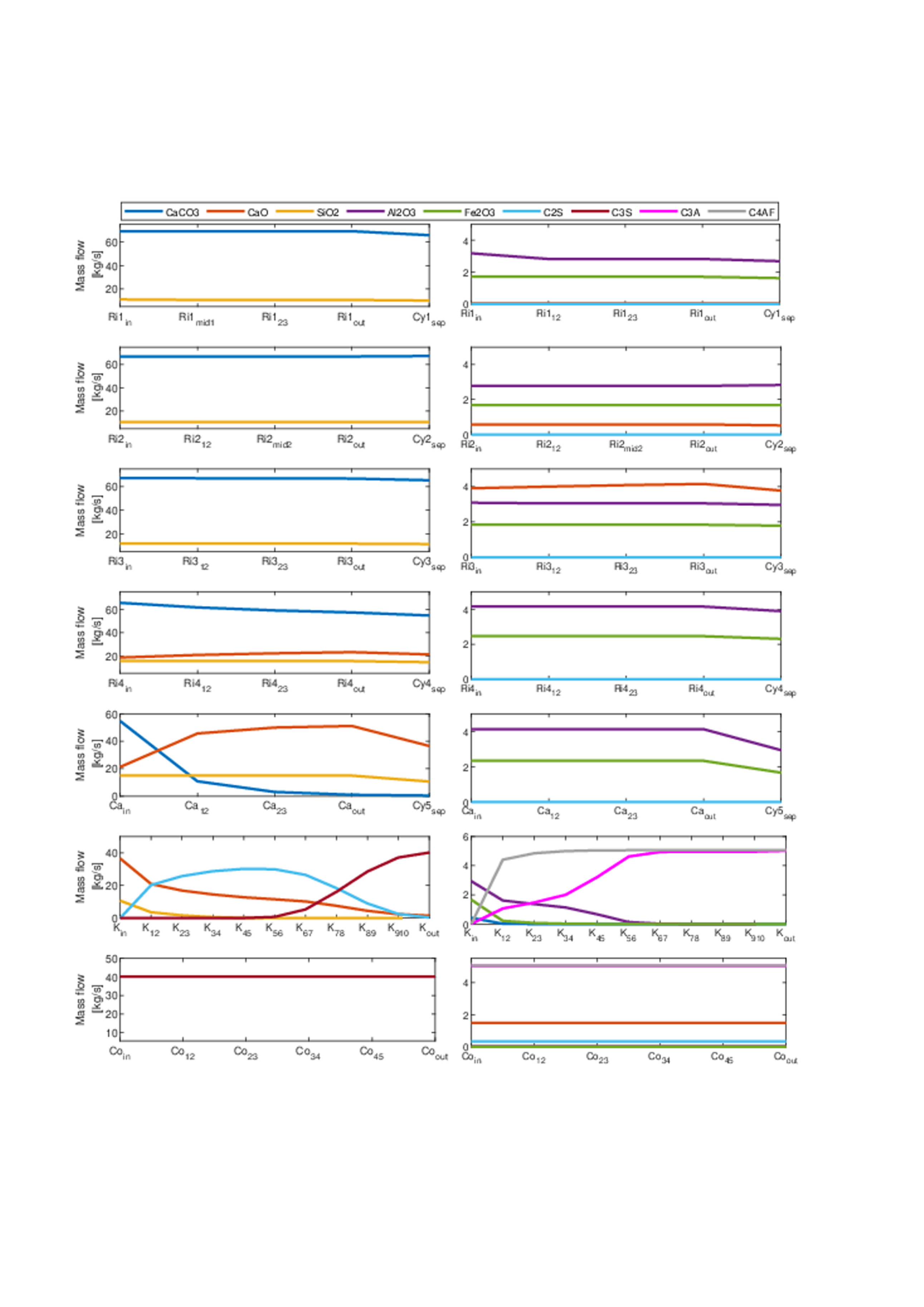}
    \caption{The steady-state mass flows of solid materials across the pyro-process. For visibility, two flow scales are used. The larger flows are shown in the left column and the smaller flows in the right column. Trace flows less than mg/s have been omitted for clarity. }
    \label{fig:MassFlow}
\end{figure*}

Fig. \ref{fig:tempKilnCooler} shows the temperature profiles of the kiln and the cooler for the end steady-state. 
We observe the clinker temperature drops rapidly upon entering the cooler from the kiln, quickly dropping below the 900$^\circ$C limit for Alite decomposition \citep{CS3toCS2}. In the kiln, the clinker temperatures are seen to rise gradually along the kiln with the gas flowing up the kiln cooling down with the profiles matching practical expectations. 
Table \ref{tab:Temp} shows specific temperatures of the kiln and the cooler for both steady-states. We can see a temperature drop of up to 21$^\circ$C between the steady-states, supporting the conclusions on the effect of the increased feed.
Extrapolating the cooler clinker temperatures, the exit temperatures of the steady states become 127.5$^\circ$C and 126.5$^\circ$C, respectively, within the 100-150$^\circ$C expected \citep{bhatty2010innovations}.
\begin{table}[t]
    \centering
    \caption{Steady-state temperatures at the beginning and the end of the kiln and cooler.}
    \begin{footnotesize}  
    \begin{tabular}{c|ccccc}\hline
 Solid  & $K_1$ & $K_{10}$ & $Co_{1}$ & $Co_{4}$& $Co_{5}$\\ \hline
 Units  & $^\circ$C & $^\circ$C & $^\circ$C & $^\circ$C & $^\circ$C\\ \hline
 start & 888.7 &  1453.0    &   686.5 &  212.8  &   155.9\\
 end & 877.0 &   1431.9 &    680.1&     211.1  &   154.7\\ \hline \hline
 Gas   & $K_1$ & $K_{10}$ & $Co_{1}$  &$Co_{4}$& $Co_{5}$ \\ \hline
 Units  & $^\circ$C & $^\circ$C & $^\circ$C & $^\circ$C & $^\circ$C\\ \hline
 start & 1177.8 &2166.4&   1059.0  &   204.3  &   139.2\\
 end  & 1167.8 & 2157.3 &  1045.7  &    203.2&     138.7\\ \hline
    \end{tabular}    
    \end{footnotesize}
    \label{tab:Temp}
\end{table} 
With the feed step, the specific heat consumption (fuel energy per mass clinker) of the kiln and calciner has dropped from 300 to 295 kcal/kg and from 675 to 664 kcal/kg respectively.

\begin{table}[t]
    \centering
    \caption{Cyclone separation efficiencies (mass \%)}
    \begin{footnotesize}  
    \begin{tabular}{c|ccccc}\hline
         &  $Cy_1$&  $Cy_2$&  $Cy_3$&  $Cy_4$&  $Cy_5$\\ \hline
         &  wt\%&  wt\%&  wt\%&  wt\%&  wt\%\\ \hline
         start &95.10 &    89.14&    87.03&    85.84&    71.34 \\
         end   &95.12 &    89.18&    87.07&    85.85&    71.53\\ \hline
         diff. &0.02 &    0.04&    0.04&    0.01&    0.19\\ \hline
    \end{tabular}    
    \end{footnotesize}
    \label{tab:cycloneEFF}
\end{table}
Fig. \ref{fig:MassFlow} shows the mass flows between each segment in the model at the end for each component.
We can see that the majority of reactions happen in the calciner and kiln module in connection with the heat generated by the combustion of fuel.
From the cyclone separations flows at the Cy$_\text{sep}$, we observe flow drops from the neighboring riser/calciner, illustrating the cyclone efficiencies.
Table \ref{tab:cycloneEFF} shows the cyclone efficiencies for the steady-states.
We can see that cyclone efficiency increases with the increased feed load, as one would expect with increased loads giving higher saltation rates. The full-system efficiencies are fairly close to the efficiencies reported in the scenario, with only cyclone 5 being more than 1\% off.


\section{Discussion}
\label{sec:Discussion}
From the simulated feed step with fixed heat supplies, we observed a small temperature decrease throughout the process.
So while clinker quality is improved by lowering the free lime content through the additional compounds, the quality worsens with respect to other content requirements, e.g., the \ce{C3S} content.
An increase in fuel intake would thus be needed to ensure the overall clinker quality, therefore, the simulator can be used to evaluate control strategies.


From Fig. \ref{fig:solidCon} and Fig. \ref{fig:TempDyn}, we observe that the system takes around 25 hours before it settles into a new steady-state configuration. This aligns well with the 20 - 48 hours it takes to start up a cold kiln. 
Thus the settling time indicates the need for control to respond quickly to changes to keep the desired performance.

Given the presented modular approach and detailed outline, other systems, e.g., dual-string preheating tower, can be constructed straightforwardly. Similarly, the system description can easily be extended to include more compounds and reactions by providing the material data and reaction rate descriptions, extending the concentration and reaction vectors. 
The model parameters have to be tuned specifically for each system; as such, data sets are a requirement to match specific real systems.
The requirement to tune the different model aspects indicates that the model can be improved further.

In the kiln, we tuned the porosity to match the fill rate to the load rate.
The introduction of a porosity or bulk density model would provide proper solid phase volumes for the kiln and cooler modules, accounting for the spatial interactions between the clinker compounds.

The tuning of the heat transfer in the kiln was introduced to sufficiently cool the gas at the kiln entrance (segment 1). The observed non-tuned temperature was around 1300$^\circ$C, instead of the tuned temperature of 1178$^\circ$C.
The practical temperature range is around 800$^\circ$C - 1100$^\circ$C \citep{bhatty2010innovations}. The formulas used for heat convection and radiation are typically empirical models. Improvements include, e.g., a theoretical model or calibration of the empirical model for the specific system. 

We observed a significant tuning of the gas velocities to correct the pressure drops.
This could indicate that a different gas velocity formulation could provide better fits than the Darcy-Weisbach equation for most modules.

Since the cyclone modules were tuned for the pressure drop, their efficiency, and solid densities, this might indicate the lumped single-cell model is too simple.
Separating the cyclone into more segments could provide a model without the need for tuning.

The reaction rates were tuned to fit the reaction to the reference performance and to account for the difference in units between those reported in the literature and applied in the model. As the models are empirical Arrhenius expressions, calibration to the specific modules or system could provide improvements. 
Other options could be shrinking core models or simply the effects of catalysts, e.g., the melt phase (\ce{C3A}, \ce{C4AF}) effects on \ce{C3S} formation.

Further development of the model could include:
\vspace{-2mm}
\begin{itemize}\setlength\itemsep{0.0em}
    \item Splitting the solid phase of the kiln into a solid phase and a melt phase.
    \item Introducing entrapped dust of solid particles in the kiln and cooler modules.
    \item Introducing degradation and build-up of the refractory phases.
    \item Addition of walls to the cooler module.
    \item Inclusion of more compounds and reactions.
\end{itemize}

%% file: Conclusion.tex
\section{Conclusion}\label{sec:Conclusion}
In this paper, we presented an index-1 DAE model for dynamic and steady-state simulation of the pyro-process in the production of clinker.
The model is formulated using a systematic modular modeling procedure based on first engineering principles involving mass and energy balances, thermo-physical properties, transport phenomena, flow pattern, geometry, stoichiometry and kinetics, and algebraic relations for the volume and the internal energy.
The full pyro-process model is the continuation of prior studies on the individual parts of the process. The paper unifies the models into a single model by outlining the interconnections, boundaries, and overall tuning for a given system.
The model is fitted to a steady-state scenario by manual calibration. 
A simulation with a step-input is provided, the simulation results match the scenario and provide a dynamic performance that qualitatively matches the expected changes for the step.
The model is an important tool for the model-based design and development of control and optimization systems for the pyro-process of cement plants, as these systems can be applied to improve the clinker quality, energy efficiency, and \ce{CO2} emissions without costly modifications of existing plants.